\documentclass[pra,twocolumn,showpacs,preprintnumbers,superscriptaddress,amsmath,tightenlines,epsfig]{revtex4}

\usepackage{bm}% bold math
\usepackage{graphicx}% Include files

\newcommand{\be}{\begin{equation}}
\newcommand{\ee}{\end{equation}}
\newcommand{\bi}{\begin{itemize}}
\newcommand{\ei}{\end{itemize}}
\newcommand{\ben}{\begin{enumerate}}
\newcommand{\een}{\end{enumerate}}

\newcommand{\rv}{{\bf r}}

\newcommand{\Bv}{{\bf B}}

\newcommand{\dv}{{\bf d}}

\newcommand{\beq}{\begin{equation}}
\newcommand{\eeq}{\end{equation}}
\newcommand{\bea}{\begin{eqnarray}}
\newcommand{\eea}{\end{eqnarray}}
\newcommand{\up}{\uparrow}
\newcommand{\down}{\downarrow}
\newcommand{\<}{\langle}
\renewcommand{\>}{\rangle}
\renewcommand{\(}{\left(}
\renewcommand{\)}{\right)}

\newcommand{\commentout}[1]{{}}
\newcommand{\half}{\hbox{$1\over2$}}
\newcommand{\eq}[1]{Eq.~\eqref{#1}}

\begin{document}
\title{Dynamical and energetic instabilities in multi-component Bose-Einstein
condensates in optical lattices}
\author{J. Ruostekoski}
\affiliation{School of Mathematics, University of Southampton,
Southampton, SO17 1BJ, UK}
\author{Zachary Dutton}
\affiliation{Naval Research Laboratory, Washington, DC 20375}

\begin{abstract}
We study dynamical and energetic instabilities in the transport
properties of Bloch waves for atomic multi-component Bose-Einstein
condensates in optical lattices in the tight-binding limit. We
obtain stability criteria analytically, as a function of superfluid
velocities and interaction parameters, in several cases for
two-component and spinor condensates. In the two-species case we
find that the presence of the other condensate component can
stabilize the superfluid flow of an otherwise unstable condensate
and that the free space dynamical miscibility condition of the two
species can be reversed by tuning the superfluid flow velocities. In
spin-1 condensates, we find the steady-state Bloch wave solutions
and characterize their stability criteria. We find generally more
regions of dynamical instability arise for the polar than for the
ferromagnetic solutions.  In the presence of magnetic Zeeman shifts,
we find a richer variety of condensate solutions and find that the
linear Zeeman shift can stabilize the superfluid flow in several
cases of interest.
\end{abstract}

\date{\today}
\maketitle

\section{Introduction}

There has been considerable recent interest in the dynamical
properties of atomic Bose-Einstein condensates (BECs) in optical
lattice potentials
\cite{AND98,BUR01,MOR01,josephson,CAT03,CRI04,FAL04,FER05,SAR05,TUC06,FER07,WU01,SME02,WU03,ZHE04,MON04}.
In optical lattices the nonlinear mean-field interaction of the BEC
may give rise to dynamical and energetic instabilities in the
transport properties of the atoms. For a single-component
condensate, when the center-of-mass (CM) velocity reaches a critical
value, the BEC dynamics become unstable resulting in an abrupt stop
of the transport of the atom cloud in the lattice. Such a superfluid
to insulator transition has a classical nature and it can be
described using the Gross-Pitaevskii (GP) mean-field models
\cite{WU01,SME02,WU03,MON04}. In the dynamically unstable regime
small initial perturbations around a moving solution grow
exponentially in time, resulting in the randomization of the
relative phases between atoms in adjacent lattice sites. The
dynamical transition to inhibited atom transport was experimentally
observed in the classical regime
\cite{BUR01,CAT03,CRI04,FAL04,SAR05} and experimental methods to
characterize both the dynamical and energetic instabilities of
moving condensates have been developed \cite{SAR05}. Inhibition of
transport was also observed in the presence of large quantum
fluctuations using strongly confined narrow atom tubes \cite{FER05}.
In a confined system with enhanced quantum fluctuations the sharp
classical transition is smeared out \cite{POL04,RUO05,BAN06},
resulting in a gradually increasing friction in the atom transport.
Due to the broadening of the velocity distribution of the atoms,
even at low velocities a non-negligible atom population occupies the
dynamically unstable high velocity region of the corresponding
classical system, generating in the shallow lattice limit the
friction \cite{RUO05,POL05}.

Despite this work on single-component condensates, there have been
relatively few studies of dynamics of multi-component BECs in
optical lattices.  Due to nearly equal trapping potentials of
different Zeeman sub-levels (for example $F=1,m_F=-1$ and
$F=2,m_F=1$ in $^{23}$Na and $^{87}$Rb), it is possible to create
long-lived two component BECs, forming an effective spin-1/2 system.
These have especially long lifetimes in $^{87}$Rb due to a
fortuitous cancelation of scattering lengths \cite{VOG02}. This
additional degree of freedom has been utilized to study an
interesting array of effects in both Bose-condensed and
non-condensed cold Bose systems, including phase separation
\cite{HALL98}, optically-induced shock waves \cite{DUT01}, spin
waves \cite{LEW02,NIK03}, overlapping $^{41}$K--$^{87}$Rb BEC
mixtures \cite{MOD02}, spin squeezing \cite{SOR01}, and vector
soliton structures \cite{RUO01,BUS01,SAV03}. Experimental work on
two-component BECs in optical lattices, from the viewpoint of
quantum logic gates, was reported in \cite{MAN03}. There has also
been a recent experimental realization a two-species
$^{41}$K--$^{87}$Rb Bose mixture in an optical lattice \cite{CAT07}.

Alternatively, in dipole traps \cite{STE98} the spin of the atom is
no longer constrained by the magnetic field and, due to the
additional atomic spin degrees of freedom, the BEC exhibits a richer
spinor order parameter structure. The spin of the optically trapped
BECs can generally have significant effects on the dynamical
properties of the BECs \cite{STE98,MIE99}, give rise to spin
textures \cite{LEA03}, and support of highly nontrivial defect
structures \cite{ZHO03,RUO03,MUE04,REI04}. Experiments have also
explored dynamics in spinor condensates in harmonic traps
\cite{SCH04,CHA04,KRO05,BLA07} and optical lattices \cite{WID05},
and the application of spinor gases to spatially resolved
magnetometry \cite{VEN07}.

In this paper we investigate both dynamical and energetic
instabilities in the transport of multi-component BECs in optical
lattices.  We first consider magnetically trapped two-component BECs
where the two condensates occupy different hyperfine states of the
same atom or are formed by mixtures of two different atoms.
Transport properties of two-component BECs in an optical lattice
were studied in Ref.~\cite{HOO06}, and numerical results for
dynamical instabilities were presented for a special case.  In
particular, dynamical instabilities were shown to arise from the
critical velocity as well as from the phase separation of the two
species. In contrast to that work, we obtain analytic results for
the condensate dynamical instability points and analyze in detail
the complete phase space of stability criteria for both dynamical
and energetic instabilities across a broad range of parameters. We
vary the intra- and inter-species interaction strengths, the site
hopping term for each spin component independently, as well as the
velocities of the two BECs, allowing application of our results to a
large variety of experimental systems.   Among the novel results
presented here are the possibility of a second BEC component
stabilizing the superfluid flow of an otherwise unstable first BEC
component (that exceeds the critical velocity of a single-component
BEC). In addition, we find the free space phase separation criteria,
that the square of the inter-species interaction coefficient exceeds
the product of the intra-species interaction coefficients
($U_{12}^2>U_{11}U_{22}$), can be reversed in an optical lattice.
This can happen if one of the BECs has a velocity larger and the
other one smaller than the single-species critical velocity (the
effective masses of the two components exhibit different signs).

We also analyze transport properties of optically trapped spin-1
BECs in optical lattices, which have not been experimentally
investigated to date.  Here we obtain analytic expressions for both
the dynamical and energetic instability regions of the Bloch wave
solutions. In contrast to the two-component case, spin changing
collisions allow the atom population of different spin components to
adjust to lower the energy of the system, according to whether the
scattering lengths correspond to polar or ferromagnetic values. Our
results illuminate the different stability properties of the polar
versus ferromagnetic solutions, which, in the absence of the Zeeman
shifts, are most apparent for large spin-dependent scattering
lengths or when the spin-dependent and spin-independent scattering
lengths exhibit different signs. The presence of the Zeeman level
shifts provides a richer variety of steady-state Bloch wave
solutions, including novel solutions that do not exist for the case
of small level shifts. We find that the quadratic Zeeman shift, due
to its role in the energy conservation of spin changing collisions,
plays an important role in the stability of various condensate
solutions. However, we also see that linear Zeeman shifts play an
important role in stabilizing many of the solutions. The dynamical
instabilities of the spinor BECs can be important, e.g., also in the
formation of solitons that have been studied in the homogeneous case
in Ref.~\cite{DAB07}.

In Sec.~\ref{twocomp} we introduce the discrete nonlinear
Schrodinger equation and the Bogoliubov-de Gennes approach to study
the stability properties of condensate solutions.  This is done in
the context of the two-component case but the same method is used
later for the spinor case.  We then derive expressions for the
normal mode energies and discuss the dynamical and energetic
stability of the two-component case.  In Sec.~\ref{spinor} we apply
this method to the spinor case. We first discuss the polar case,
then the ferromagnetic case, then finally the effect of the Zeeman
shifts. Some experimental considerations for observation of the
effects studied are discussed in Sec.~\ref{experimental}.  We
summarize our results in Sec.~\ref{conclusion}. The Bogoliubov-de
Gennes matrices are presented explicitly in Appendix~\ref{app} and
the detailed analysis of the stability of the two-component case
when the phase separation condition is reversed is given in
Appendix~\ref{appdynamical}.

\section{A two-component condensate in an optical lattice} \label{twocomp}
\subsection{Two species system description}

Two-component BECs can be prepared in magnetic traps by
simultaneously confining different atomic species in the same trap.
The atoms may occupy two different hyperfine states of the same
atomic species or form a mixture of two condensates of two different
atomic species. For instance, two BEC components in perfectly
overlapping isotropic magnetic trapping potentials were
experimentally realized in hyperfine spin states of $^{87}$Rb,
$|\up\>\equiv|F=2, m_f =1 \rangle$ and $|\down\>\equiv |F=1,m_f=-1
\rangle$. In this system the inter- ($a_{\up\down}$) and
intraspecies ($a_{\up\up}$ and $a_{\down\down}$) interaction
strengths are nearly equal, with
$a_{\down\down}:a_{\up\down}:a_{\up\up}::1.024:1:0.973$
\cite{HAR02}. Since the scattering lengths satisfy
$a_{\up\down}^2\agt a_{\up\up}a_{\down\down}$, the two species
experience dynamical phase separation and can strongly repel each
other \cite{HALL98}.  A more strongly repelling two-component system
of different species was created using a $^{41}$K--$^{87}$Rb mixture
\cite{MOD02}. The interatomic interactions of two magnetically
trapped BEC components do not mix the atom population and the atom
numbers of both species are separately conserved.

The dynamics of the BECs follow from the coupled Gross-Pitaevskii
equation (GPE)
\begin{equation}
i \hbar {\partial \psi_j\over \partial t} = (
-{\frac{\hbar^2}{2m_j}}{\bf \nabla}^2+V_j(\rv) +\sum_k \kappa_{jk}
|\psi_k|^2 ) \psi_j\,. \label{gpe}
\end{equation}
Here we have defined the interaction coefficients $\kappa_{ii}\equiv
4\pi\hbar^2 a_{ii}/m_i$ and $\kappa_{ij}\equiv 2\pi\hbar^2
a_{ij}/\mu$ ($i\neq j$), where the wavefunctions are normalized to
$N_j$, $N=N_1+N_2$ is the total atom number, $m_j$ is the atomic
mass of BEC component $|j\>$, and $\mu=m_1m_2/(m_1+m_2)$ is the
reduced mass. The intraspecies and the inter species scattering
lengths are denoted by $a_{ii}$ and $a_{ij}$ ($i\neq j$),
respectively. The external potential is generally a superposition of
a harmonic trapping potential
$V_H^{(j)}(\rv)=m(\omega_{jx}^2x^2+\omega_{jy}^2y^2+\omega_{jz}^2z^2)/2$
and the periodic optical lattice potential $V_L^{(j)}(\rv)=V_0^{(j)}
\sin^2(\pi x/a +\varphi_j)$,
$V_j(\rv)=V_H^{(j)}(\rv)+V_L^{(j)}(\rv)$, where $a$ denotes the
lattice spacing. In the following we ignore the effect of the
harmonic trapping potential along the lattice and consider the
system as translationally invariant. We also neglect density
fluctuations orthogonal to the optical lattice and consider the
dynamics as effectively 1D.

\begin{figure}
\includegraphics[width=0.8\columnwidth]{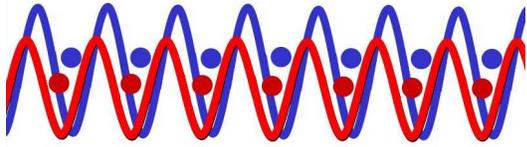}\vspace{-0.2cm}
\caption{The two BECs may experience different optical lattice
potentials that can be shifted with respect to each other. The
interspecies interaction strength $U_{12}$ is proportional to the
overlap integral of the two lattice site mode functions and it can
be adjusted by shifting the lattices. Moreover, the kinetic energy
hopping amplitude of the two species $J_1,J_2$ may be independently
modified by changing the barrier height between the neighboring
sites.} \label{latticefig}
\end{figure}

We write the GPE in the tight-binding approximation by expanding the
BEC wavefunctions on the basis of the Wannier functions and only
keep the lowest vibrational states in each lattice site $\eta$, so
that $\psi_j(\rv)=\sum_\eta c^{(j)}_{\eta}\phi_{j\eta}(\rv)$
\cite{JAK98}. We obtain discrete nonlinear Schr\"odinger equations
(DNLSEs):
\begin{equation}
i \hbar {d c^{(j)}_{\eta}\over d t} =
-J_j(c^{(j)}_{\eta+1}+c^{(j)}_{\eta-1})
 +\sum_k U_{jk}
|c^{(k)}_{\eta}|^2  c^{(j)}_{\eta}\,. \label{gpe2}
\end{equation}
With similar assumptions, we have the hopping amplitude, $J_j>0$,
for the atoms between adjacent lattice sites:
\beq
J_j\simeq-\int
d^3r\,({\hbar^2\over 2m_j}\nabla
\phi_{j\eta}^*\cdot\nabla\phi_{j,\eta+1}+\phi_{j\eta}^* V_L^{(j)}
\phi_{j,\eta+1})\,. \label{J}
\eeq
The nonlinearities are given by $U_{jk}\simeq\kappa_{jk}\int
d^3r\,|\phi_{j\eta}|^2 |\phi_{k\eta}|^2$.  The two BEC species may
generally experience different lattice potentials \cite{MAN03} and
so the interspecies coupling coefficient $U_{12}$ may be varied by
displacing the two lattice potentials with respect to each other (to
modify the overlap integral between the wavefunctions), even when
the values of the scattering lengths remain constant;
Fig.~\ref{latticefig}.

\subsection{Dynamical stability of two species}\label{2exc}

\subsubsection{Collective two-component excitations}

We study the stability of plane wave solutions to \eq{gpe2} by
investigating the effect of small perturbations around the carrier
wave. Our treatment is analogous to the approach in
Ref.~\cite{SME02} to analyze single-component BECs. For the constant
atom density along the lattice, the Bloch waves
$c^{(j)}_{\eta}=\sqrt{n_j} \exp{[i(k_j a \eta-\nu_j t)]}$ that
satisfy \eq{gpe2} exhibit the frequency $\nu_j=\sum_i n_i
U_{ji}-2J_j\cos(k_ja)$, where $n_j=|c^{(j)}_{\eta}|^2$ denotes the
constant atom population in spin state $j$ at each site. The
perturbed carrier wave can be written as a Bogoliubov expansion:
\begin{equation} \label{BogAnsatz}
 c^{(j)}_{\eta}=(\sqrt{n_j} +u_j e^{iqa\eta}-v_j^*e^{-iqa\eta})
e^{i(k_j a \eta-\nu_j t)}\,.
\end{equation}
Here the $k_j$ represent the potentially non-zero velocities of the
condensate in each component $j$.  The stability of such states in a
single component BEC has been studied experimentally in
Refs.~\cite{FAL04,SAR05} by dynamically moving the lattice
potential. Substituting this expansion into \eq{gpe2} and
linearizing the equations in $u_1$, $v_1$, $u_2$, and $v_2$ yields a
system of four equations
\begin{equation} \label{BogDeGennes2}
i{d\over dt} {\bf x} = \sigma {\cal M}(q) {\bf x},\quad {\bf
x}=\begin{pmatrix}
u_1 \\
v_1\\
u_2\\
v_2
\end{pmatrix},\quad \sigma=\left(
\begin{array}{cc}
\sigma_z & 0 \\
0 & \sigma_z
\end{array}
\right) \,,\end{equation}
where $\sigma_z$ denotes the $2\times2$ Pauli spin matrix. The
elements of the $4\times4$ matrix ${\cal M}(q)$ follow from the
linearization procedure as in the single-component BEC case.  We
write out this matrix explicitly in Appendix \ref{app}. The
eigenvalue problem for the matrix $\sigma {\cal M}(q)$ can be solved
analytically. In order to preserve the symmetry properties the
Bogoliubov equations and to obtain simple analytic expressions for
the normal mode energies we require that the atom currents of the
two BECs are equal, i.e., $J_1\sin(k_1 a)=J_2\sin(k_2 a)$.

The eigenvalues represent the normal mode energies and read
\begin{widetext}
\beq
\omega_q=2J_1 \sin (k_1 a) \sin (q a)\pm
\sqrt{\half(\omega_{1,q}^2+\omega_{2,q}^2)\pm
\half\sqrt{(\omega_{1,q}^2-\omega_{2,q}^2)^2+16\epsilon_{1,q}\cos
(k_1a)\epsilon_{2,q}\cos(k_2a)n_1n_2U_{12}^2}}\,.\label{2becf}
\eeq
\end{widetext}
The four eigenvalues correspond to all permutations of the $\pm$
signs. Only two of the eigenvalues are independent.  The first term
in \eq{2becf} represents the Doppler shift of the excitation
energies due to the superfluid current. Here $\omega_{j,q}$ denotes
the single-condensate normal mode energies (without the Doppler
shift term)
\beq
\omega_{j,q}^2=\epsilon_{j,q}\cos(k_ja)[\epsilon_{j,q}\cos(k_ja)+2
n_j U_{jj}]\,,\label{1bec}
\eeq
and
\beq
\epsilon_{j,q}=4J_j\sin^2\({q a\over2}\)\,, \label{ideal}
\eeq
is the spectrum of an ideal, non-moving BEC.

For the case of positive definite ${\cal M}(q)$ all the eigenvalues
$\omega_q$ of $\sigma{\cal M}(q)$ are real. In that case the
physical solutions of the corresponding eigenvectors ${\bf y}$
exhibit positive normalization ${\bf y}^\dagger \sigma {\bf y}=1$
(the `$+$' sign in the front of the first square root) and
unphysical eigenvectors negative normalization ${\bf y}^\dagger
\sigma {\bf y}=-1$ (the `$-$' sign in the front of the first square
root). The eigenvalues $\omega_q$ with a nonvanishing imaginary part
are associated with eigenvectors satisfying ${\bf y}^\dagger \sigma
{\bf y}=0$.  The BEC system becomes dynamically unstable when the
normal mode frequencies in \eq{2becf} exhibit nonvanishing imaginary
parts, indicating perturbations that grow exponentially in time.
Such modulational instabilities occur in a closed system due to the
nonlinear dynamics and do not require energy dissipation. The rate
at which the instability sets in depends on the magnitude of the
imaginary part of the eigenfrequency.

For small momenta, $q a\ll 1$, $\epsilon_{j,q}\simeq J_jq^2
a^2=\hbar^2 q^2/2m_j^*$, where we introduced the effective mass of a
noninteracting BEC as $m_j^*=\hbar^2/(2J_j a^2)$. Similarly, we
obtain $2J_1 \sin (k_1 a) \sin (q a)\simeq 2 J_1 k_1 q a^2 = \hbar^2
k_1 q/m_1^*$ reinforcing the interpretation of the first term in
\eq{2becf} as the Doppler shift contribution.

If we set $U_{12}=0$ in \eq{2becf}, we obtain independently the
decoupled normal mode energies of the two BECs $\omega_q=2J_1 \sin
(k_1 a) \sin (q a)+ |\omega_{1,q}|$ and $\omega_q=2J_1 \sin (k_1 a)
\sin (q a)+ |\omega_{2,q}|$, analogously to the single-condensate
normal modes obtained in Ref.~\cite{SME02}.

The intraspecies interaction $U_{12}$ mixes the normal modes of the
two BECs. In the experimentally interesting regime $n_jU_{jj}\gg
J_j$, if $U_{12}^2\simeq U_{11}U_{22}$, one of the frequencies
approaches zero indicating an instability similar to the uniform
two-component BEC system. Specifically, for $k_1=k_2=0$, we obtain
in that case $\omega_{q,+}^2\simeq \omega_{1,q}^2+\omega_{2,q}^2$
and $\omega_{q,-}^2\ll \omega_{1,q}^2,\omega_{2,q}^2$.

By expanding $\omega_q$ for small $q$ in \eq{2becf} with
$k_1=k_2=0$, we obtain $\omega_q\simeq \hbar s q$ where $s$ is the
speed of sound
\begin{widetext}
\beq \label{s} s_\pm={a\over
\hbar}\sqrt{J_1n_1U_{11}+J_2n_2U_{22}\pm
\sqrt{(J_1n_1U_{11}-J_2n_2U_{22})^2+4 J_1 J_2n_1n_2 U_{12}^2}}\,.
\eeq
\end{widetext}
The long wavelength excitations are unstable when one of the
solutions for the speed of sound has an imaginary part.

If we do not assume that the two BEC currents are equal $J_1\sin(k_1
a)\neq J_2\sin(k_2 a)$, the Doppler shifts for the two BECs are
different and the usual symmetry properties between the positive and
the negative energy Bogoliubov eigenfunctions are lost. We still
find analytic solutions for the eigenenergies $\omega_q$, but these
no longer have simple compact expressions as in \eq{2becf}. The
basic formalism may be used to generate stability diagrams in these
cases numerically, for instance, even if the two BECs have the
velocities in the opposite directions. In the following we
concentrate on analyzing the general features of the two-component
system that may already be obtained from \eq{2becf}.

\subsubsection{Stability with equal signs for $\cos(k_1a)$ and $\cos(k_2a)$}
\label{samesigns}

We first analyze the dynamical stability of the two-component
system, given in \eq{2becf} for $J_1\sin(k_1 a)=J_2\sin(k_2 a)$, for
the case that $\cos(k_1a)$ and $\cos(k_2a)$ exhibit equal sign.  For
that case, the expression inside the square root in \eq{2becf} is
always negative (for any values of $U_{12}$), and the dynamics
unstable, if $\omega_{1,q}^2+\omega_{2,q}^2<0$. Physically, this
corresponds to the situation where the two-component dynamical
instability is driven by the instabilities of the individual
single-component BEC excitations \eqref{1bec}. We have
$\omega_{1,q}^2+\omega_{2,q}^2<0$, for some values of $q$, if
\beq
\sin^2{\({q a\over 2}\)} < -{
D_{11}+D_{22}\over 2[J_1^2\cos^2{(k_1a)}+J_2^2\cos^2{(k_2a)}]}\,,
\eeq
where
\beq
D_{ij}=J_i\cos(k_ia)n_jU_{jj}\,.\label{Dco}
\eeq

The inequality is most easily satisfied for excitations in the
long-wavelength limit $q\rightarrow0$, and is satisfied when the
right hand side is positive, i.e., for
\beq
D_{11}+D_{22}<0\,.
\label{insta}
\eeq
According to \eq{insta}, the modes can become unstable if
$k_1a,k_2a>\pi/2$ even when $U_{11},U_{22}>0$, which is the usual
high velocity instability seen in the single component case
\cite{WU01}.   The dynamics can also be unstable when
$k_1a,k_2a<\pi/2$, for negative values of $U_{11}$ and $U_{22}$.

In addition to the instability occurring for
$\omega_{1,q}^2+\omega_{2,q}^2<0$, the two-component system in
\eq{2becf} is always dynamically unstable (for the case that
$\cos(k_1a)$ and $\cos(k_2a)$ exhibit equal sign) if
\beq
U_{12}^2>U_{11}U_{22}\,, \label{phasesep}
\eeq
as the expression in the outer square root in \eq{2becf} becomes
negative. In particular, the unstable $q$ modes are those that
satisfy (when $\omega_{1,q}^2+\omega_{2,q}^2\geq0$):
\begin{widetext}
\beq
\sin^2{\({qa\over 2}\)} < {-(D_{12}+D_{21})+\sqrt{(D_{12}-D_{21})^2
+4 J_1\cos(k_1a)J_2\cos(k_2a) n_1 n_2 U_{12}^2}\over 4
J_1\cos(k_1a)J_2\cos(k_2a)}\,.
\eeq
\end{widetext}
The two-component system, consequently, is in this case dynamically
stable if $D_{11}+D_{22}> 0$ and $U_{12}^2\leq U_{11}U_{22}$.

The unstable dynamics for $U_{12}^2>U_{11}U_{22}$ correspond to the
analogous instability which occurs in the free space case due to
phase separation. One should emphasize, however, that the value of
$U_{12}$ is not only determined by the inter-species scattering
length, but also by the spatial overlap integral of the lattice site
wavefunctions for the two species; see the definition of $U_{ij}$
below Eq.~(\ref{J}). By means of shifting the relative position of
the two BEC lattice potentials, one may easily reduce the value of
$U_{12}$.

\begin{figure}
\includegraphics[width=0.95\columnwidth]{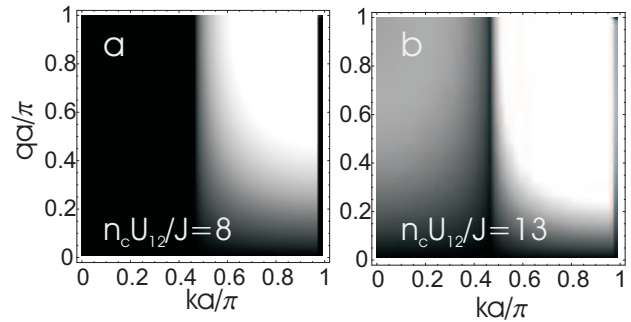}
\caption{\label{fig:twocompKQ} \textbf{(a)} Largest imaginary part
of the eigenvalues Eq.~(\ref{2becf}) as function of $k \equiv
k_1=k_2$ and $q$ for the case $n_1 U_{11}=n_2 U_{22}$, $n_c U_{12}=8
J_1$ and $J_2=J_1$. We define $n_c \equiv \sqrt{n_1 n_2}$. Gray
scale goes from $0$ (black) to $5 J_1$ (white). \textbf{(b)} Same
plot with $n_c U_{12}=13 J_1$ (phase separation regime). }
\end{figure}

Phase space diagrams of the dynamical instability strengths as a
function of the excitation wavelength $q$ and condensate wave number
$k$ are shown in Fig.~\ref{fig:twocompKQ}. In
Fig.~\ref{fig:twocompKQ}(a) we see the system is stable for the case
$n_c U_{12}=8 J_1$, $n_1 U_{11}=n_2 U_{22}=10 J_1$ (defining
$n_c=\sqrt{n_1 n_2}$) when $ka<0.5 \pi$, then becomes unstable for
$ka>0.5 \pi$, in accordance with \eq{insta}. The strength of the
instabilities (the largest imaginary part of the eigenvalues) are
linear in $q$ [for the slope, see Eq.~(\ref{s})] until they saturate
approximately at a value $ {\rm Im}(\omega_q)\simeq \sqrt{-16 n_c
U_{12} J_1 \cos{(k_1a)}}$, which is $\approx 11 J_1$ at $ka=\pi$ in
Fig~\ref{fig:twocompKQ}(a). The strongest instability at $ka=\pi$
represents period doubling that drives the system away from the
Bloch state. The behavior is qualitatively similar for all $n_c
U_{12} \le n_1 U_{11}=n_2 U_{22}$ and also for smaller values of
$J_1$ and $J_2$. However, for $n_c U_{12}> n_1 U_{11}$
[Fig.~\ref{fig:twocompKQ}(b)], we see the predicted phase separation
instability arises in the $ka<0.5 \pi$ region. It is interesting to
note this instability is markedly weaker than the high velocity
instability ($ka>0.5 \pi$), as its maximum value (occurring for
$k=0$, $q a=\pi$) scales as ${\rm Im}(\omega_q)\simeq \sqrt{8 (n_c
U_{12}-n_1 U_{11}) J_1}$ (in the limit that $n_c U_{12}-n_1 U_{11}
\gg J_1$). In the case plotted in Fig.~\ref{fig:twocompKQ}(b), the
value saturates at $\approx 3 J_1$, whereas the instability strength
in the $k a >0.5 \pi$ region reaches $\approx 13 J_1$.

\begin{figure}
\includegraphics[width=0.95\columnwidth]{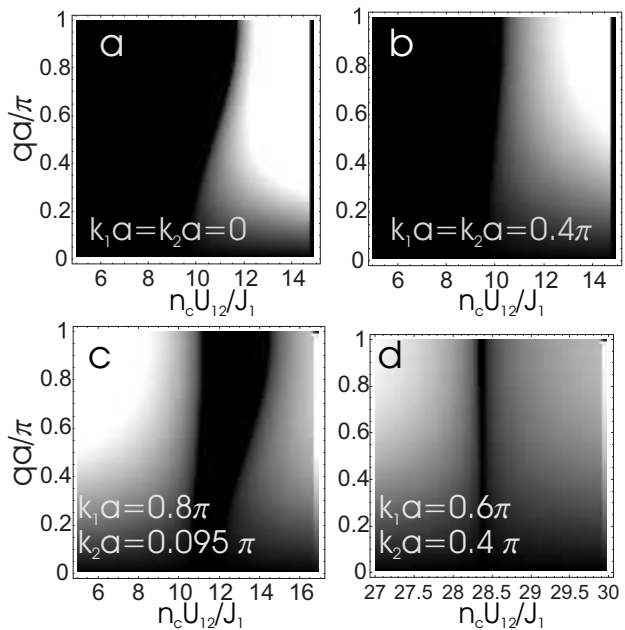}
\caption{\label{fig:twocompU12Q} \textbf{(a)-(b)} Imaginary part of
the eigenvalues versus $n_c U_{12}$ (again $n_c=\sqrt{n_1 n_2}$),
holding $n_1 U_{11}=n_2 U_{22}=10 J_1$ for the indicated condensate
velocities $k$.  Here $J_2=J_1$ \textbf{(c)} A case with opposite
signs of $\cos(k_1 a)$ and $\cos(k_2 a)$, showing a region of
dynamical stability for all $q$. Here $n_1 U_{11}=10 J_1$, $n_2
U_{22}=12 J_1$ and $J_2=2 J_1$. A similar narrow stable region
exists around $n_c U_{12} \simeq -12 J_1$. \textbf{(d)} Another case
with opposite signs of $\cos(k_1 a)$ and $\cos(k_2 a)$. Here $n_1
U_{11}=27 J_1$, $n_2 U_{22}=30 J_1$ and $J_2=J_1$. Again another
stable region in this case is located close to $n_c U_{12} \simeq
-28.5 J_1$.  In these plots the gray scale goes from 0 to $5
J_{1}$.}
\end{figure}

\begin{figure}
\includegraphics[width=0.95\columnwidth]{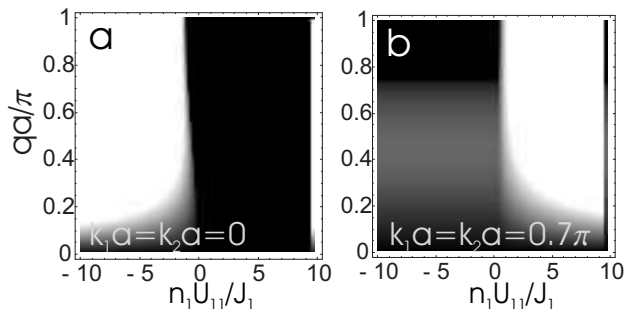}
\caption{\label{fig:twocompUQ} Imaginary part of the eigenvalues
versus $q$ and varying $n_1 U_{11}=n_2 U_{22}$, keeping $n_c
U_{12}=n_1 U_{11}-J_1$ and with $J_2=J_1$. In this figure, the gray
scale runs from 0 to $2.5 J_1$.}
\end{figure}

Figures~\ref{fig:twocompU12Q}(a,b) compare the phase separation
instability versus $n_c U_{12}$ for $k_1 a=k_2 a =0$ and $k_1 a=k_2
a= 0.4 \pi$.  Generally speaking, the largest imaginary value
reaches a maximum value ${\rm Im}(\omega_q)\simeq \sqrt{8(n_c
U_{12}-n_1 U_{11})J_1}$. The dependence on the sign of
$U_{11},U_{22}$ is shown in Fig.~\ref{fig:twocompUQ}.  We see in
Fig.~\ref{fig:twocompUQ}(a) that, for  $n_c U_{12}<n_1 U_{11}=n_2
U_{22}$ and $\cos (k_1 a)>0$, the instability is restricted to the
attractive cases $U_{11}<0$. Fig.~\ref{fig:twocompUQ}(b)
demonstrates this instability switches to the repulsive case for
$\cos(k_1 a)<0$.  Again these instabilities reach strengths ${\rm
Im}(\omega_q)\simeq \sqrt{16 n_1 U_{11} J_1}$.  However, for the
$\cos(k_1 a)<0$ case there occurs also a much weaker instability for
attractive interactions with strength ${\rm Im}(\omega_q)\simeq
\sqrt{8 (n_c U_{12}-n_1 U_{11})J_1}$.

\subsubsection{Stability with different signs for $\cos(k_1a)$ and $\cos(k_2a)$}
\label{diffsigns}

The case that $\cos{k_1a}$ and $\cos{k_2a}$ have different signs
represents a configuration where the velocities of the two BECs are
located on the opposite sides of the deflection point in the ideal,
single-particle BEC excitation spectrum \eqref{ideal} (the effective
masses of the two components exhibit different signs) and is
presented in detail in Appendix \ref{appdynamical}. In that
situation, we also always find a dynamical instability when
$\omega_{1,q}^2+\omega_{2,q}^2<0$, resulting in the relation similar
to \eq{insta}. In this case, however, the high velocity instability
condition is highly nontrivial, depending on the values of the
hopping amplitudes, the interaction strengths, atom numbers, and the
velocities: One of the BECs that reaches the single-component
critical velocity $k_1a>\pi/2$ may, or may not, destabilize the
two-component BEC system, depending on the parameter values.
Moreover, for $\omega_{1,q}^2+\omega_{2,q}^2\geq 0$, the dynamical
stability condition due to the phase separation of the non-moving
system, $U_{12}^2<U_{11}U_{22}$, is reversed, so that the {\it
entire} dynamically stable region occurs, when
$U_{12}^2>U_{11}U_{22}$. In particular, we find in that case the
system to be dynamically stable if $U_{12}$ satisfies, depending on
the value of $U_{11}$, either $U_{11}U_{22}<U_{12}^2 < \xi_1$ or
$U_{11}U_{22}< \xi_2<U_{12}^2 < \xi_1$, where $\xi_1$ and $\xi_2$
are defined in Eqs.~\eqref{xi1} and~\eqref{xi2}. Interestingly, this
also represents a situation where the other condensate component can
stabilize the superfluid flow of an otherwise unstable condensate
(exceeding the single-component critical velocity). Moreover, the
two-component system may be dynamically stable even for $U_{11}<0$
and $U_{22}>0$, since in that case $U_{11}U_{22}<U_{12}^2$.

Figures~\ref{fig:twocompU12Q}(c-d) show cases with $\cos (k_1 a)$
and $\cos(k_2 a)$ of different sign [but satisfying the condition
$J_1 \sin (k_1 a)=J_2 \sin (k_2 a)$]. For the parameters of
Fig.~\ref{fig:twocompU12Q}(c) there is a range of $n_c U_{12}$ for
which the system is dynamically stable. In this case $n_1 U_{11}$
satisfies \eq{u11b} and the dynamically stable region, according to
\eq{insta5}, is $\xi_2< U_{12}^2 < \xi_1$. The relevant quantities
are $n_c \sqrt{\zeta_2}=11.5 J_1$ and $n_c \sqrt{ \zeta_1}=12.5 J_1$
for this example. Note that the dynamical instabilities tend to be
much weaker on the large $U_{12}$ side of this stability range. In
Fig.~\ref{fig:twocompU12Q}(c) we have different hopping amplitudes
$J_1 \not= J_2$. Fig.~\ref{fig:twocompU12Q}(d) shows a case with a
smaller range of stable $n_c U_{12}$ but with $J_1 = J_2$. In this
case $n_1 U_{11}$ satisfies \eq{u11a} and the dynamically stable
region, according to \eq{insta4}, is for $U_{11}U_{22}< U_{12}^2 <
\xi_1$. Here $n_c \sqrt{U_{11}U_{22}}=28.46 J_1$ and $n_c \sqrt{
\zeta_1}=28.50 J_1$.

\subsection{Energetic stability}

The energetic stability of the superfluid flow of the homogeneous
two-component mixture depends on the properties of the energy
functional. The second-order variations of the energy for small
perturbations in the carrier wave are determined by the matrix
${\cal M}(q)$ and the system is energetically stable if ${\cal
M}(q)$ is positive definite. If any of the eigenvalues of ${\cal
M}(q)$ are negative, the system may relax to a state with lower
energy by means of dissipative coupling to the environment. The rate
at which such relaxation happens depends on the strength of the
coupling, e.g., on the number of thermal atoms interacting with the
condensate.

\begin{figure}
\includegraphics[width=0.95\columnwidth]{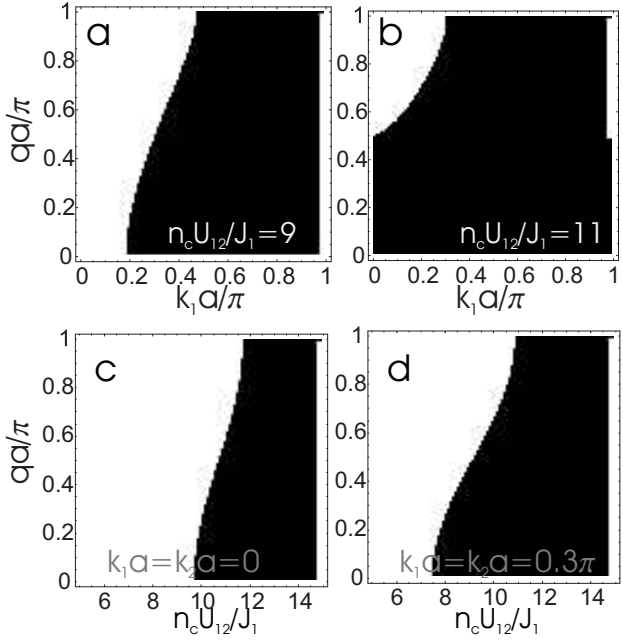}
\caption{\label{fig:twocompKQenergy} \textbf{(a)-(b)} Regions of
energetic stability (white) and instability (black) versus $k_1$ and
$q$ for the cases indicated.  We choose $n_1 U_{11}=n_2 U_{22}=10
J_1$, $J_2=J_1$ and $k_2=k_1$. \textbf{(c)-(d)} Energetic stability
regions versus $n_c U_{12}$ for a stationary and moving condensate
case.}
\end{figure}

The eigenvalues of ${\cal M}(q)$ can also be evaluated analytically
but the full solutions are rather lengthy. In
Fig.~\ref{fig:twocompKQenergy} we show the energetically unstable
regions of the two-component dynamics. Note that positive definite
${\cal M}(q)$ implies real eigenvalues $\omega_q$ of $\sigma {\cal
M}(q)$ in \eq{2becf}, so the dynamically unstable region always
forms a subset of the energetically unstable region.
Figures~\ref{fig:twocompKQenergy}(a-b) show cases with $n_c U_{12}<
n_1 U_{11}$ and $n_c U_{12}>n_1 U_{11}$, respectively.  In the
latter case there are regions of instability for some $q$ at all
condensate wavenumbers $k$.  In the former case, there are is a band
of $k$ with all $q$ modes energetically stable, with a width
proportional to $\sqrt{2J_1(n_1 U_{11}-n_c U_{12})}$, which
determines the speed of sound for the spin wave; see \eq{s} with
$n_1 U_{11}=n_2 U_{22}$, $J_1=J_2$. This is analogous to the
single-component case \cite{WU01} where there is a band of
energetically stable $k$ with a width proportional to the speed of
sound $\propto \sqrt{2J_1 n_1 U_{11}}$. In
Figs.~\ref{fig:twocompKQenergy}(c),(d) we show the dependence on
$U_{12}$.  It is seen that at $k=0$ instability only occurs for
$U_{12}>U_{11}$ while for finite $k$ the condition for stability
becomes more stringent.  For $ka>0.5 \pi$ the entire region is
unstable.  We also calculated that the specific parameter regimes
corresponding to dynamical stability with two different condensate
wavenumbers $k_1,k_2$ (Figs.~\ref{fig:twocompU12Q}(c,d)) are
energetically unstable at all $q$.

\section{A spin-1 condensate in an optical lattice} \label{spinor}

\subsection{Spinor Gross-Pitaevskii equations}

We now consider a BEC of spin-1 atoms. In the absence of a magnetic
trapping potential, the macroscopic BEC wave function is determined
by a spinor wave function $\Psi $ with three complex components
\cite{Pethick}. The Hamiltonian density of the classical GP
mean-field theory for this system reads:
\begin{align} \label{hamiltonian} {\cal H} =& \frac{\hbar^2}{2m} |\nabla
\Psi|^2 + V \rho + {c_0\over2} \rho^2 + {c_2\rho^2\over2} |\< {\bf
F}\> |^2 \nonumber\\ &+g_1 \< \Bv\cdot {\bf F}\> \rho +g_2 \<
(\Bv\cdot {\bf F})^2\> \rho \, ,
\end{align}
In Eq.~(\ref{hamiltonian}), ${\bf F}$ is the vector formed by the
three components of the $3\times3$ Pauli spin-1 matrices \cite
{Pethick}, $\< {\bf F} \>= \Psi ^{\dagger}\cdot {\bf F}\cdot
\Psi/\rho$ denotes the average spin, and $\rho(\rv)=|\Psi(\rv)|^2$
the total atom density. The weak external magnetic field is denoted
by $\Bv$ and is assumed to point along the $z$ axis. The magnetic
field produces the linear and quadratic Zeeman level shifts whose
effect is described by the last two terms in \eq{hamiltonian}. As in
the two-component case, the external potential $V$ is the sum of the
harmonic part $V_H$ (in this case due to the optical dipole trap)
and the optical lattice potential $V_L$: $V(\rv)=V_H(\rv)+V_L(\rv)$.
In the following we ignore the harmonic potential and, for
simplicity, assume that the lattice potential is the same for all
the spinor components.   Thus the Wannier basis functions
$\phi_\eta$ of our discrete basis no longer depend on the internal
state $j$. Here $c_0$ and $c_2$ are the spin-independent and
spin-dependent two-body interaction coefficients. In terms of the
s-wave scattering lengths $a_0$ and $a_2$, for the channels with
total angular momentum zero and two, they are: $c_0\equiv
4\pi\hbar^2(2a_2+a_0)/3m$, and $c_2\equiv 4\pi\hbar^2(a_2-a_0)/3m$.
For $^{23}$Na, $(a_2-a_0)/3\simeq 2a_B$ and $(2a_2+a_0)/3\simeq
50a_B$, where $a_B = 0.0529$~nm is the Bohr radius \cite{Pethick},
indicating $c_2/c_0\simeq 0.04$. In contrast to the two-component
case, in which the atom number in the two-components do not mix, in
the spinor case a homogenous condensate wavefunction can adjust
itself by varying the relative atom populations. From
Eq.~(\ref{hamiltonian}) in the absence of the external magnetic
field we immediately observe that, since $c_2>0$ for $^{23}$Na,
corresponding to the {\em polar} phase, the energy is minimized by
setting $\langle {\bf F} \rangle ={\bf 0}$ throughout the BEC for
the case of a uniform order parameter field. Alternatively, for
$^{87}$Rb we have $c_2/c_0\simeq -0.0036$ \cite{WID06}. The
parameter values for $^{87}$Rb correspond to the {\em ferromagnetic}
phase, since $c_2 < 0$, and the energy in Eq.~(\ref{hamiltonian}) in
the absence of the external magnetic field is minimized when $|
\langle {\bf F} \rangle |=1$ throughout the BEC for the case of a
uniform spin distribution.

Again using the lowest band of the Wannier state basis, the DNLSEs
are written:
\begin{align}\label{GPspinor}
i \hbar \frac{\partial c^{(+)}_\eta}{\partial t} & =  - J
    (c^{(+)}_{\eta+1} + c^{(+)}_{\eta-1})+ \delta_+ c^{(+)}_{\eta} \nonumber \\
     &  + U_0
    \sum_{\alpha=+,0,-}
    |c^{(\alpha)}_\eta|^2 c^{(+)}_\eta   \nonumber \\
    &  + U_2
    (|c^{(+)}_\eta|^2-|c^{(-)}_\eta|^2+|c^{(0)}_\eta|^2)c^{(+)}_\eta
    \nonumber \\
    &  + U_2 c^{(-)}_\eta {}^* c^{(0)}_\eta{}^2 \nonumber \\
i \hbar \frac{\partial c^{(-)}_\eta}{\partial t} & = - J
    (c^{(-)}_{\eta+1} + c^{(-)}_{\eta-1}) + \delta_- c^{(-)}_{\eta}\nonumber \\
    & + U_0 \sum_{\alpha=+,0,-}
    |c^{(\alpha)}_\eta|^2 c^{(-)}_\eta  \nonumber \\
    & + U_2
    (|c^{(-)}_\eta
    |^2-|c^{(+)}_\eta|^2+|c^{(0)}_\eta|^2)c^{(-)}_\eta
    \nonumber \\
    & + U_2 c^{(+)}_\eta{}^* c^{(0)}_\eta{}^2 \nonumber \\
i \hbar \frac{\partial c^{(0)}_\eta}{\partial t} & =  - J
    (c^{(0)}_{\eta+1} + c^{(0)}_{\eta-1})  \nonumber \\
    &  + U_0 \sum_{\alpha=+,0,-}
    |c^{(\alpha)}_\eta|^2 c^{(0)}_\eta \nonumber \\
    &  + U_2
    (|c^{(+)}_\eta|^2+|c^{(-)}_\eta|^2)c^{(0)}_\eta
   \nonumber \\
    & + 2 U_2 c^{(+)}_\eta c^{(-)}_\eta c^{(0)}_\eta{}^*
\end{align}
\noindent  Here $J$ is defined as before (\ref{J}), but with no
dependence on the internal state $j$ \cite{comment} and
$U_{0,2}\simeq c_{0,2}\int d^3r\,|\phi_{\eta}|^4$. The primary
qualitative difference with the two-component case, seen in the last
term of each of these equations, is the allowance of spin exchange
collisions.  The additional energy shift terms  $\delta_+,\delta_-$
account for Zeeman shifts (with respect to the level $m_F=0$) due to
an external magnetic field $\mathbf{B}$. For simplicity, we ignore
any effects of magnetic field gradients.

We study the stability of moving Bloch wave solutions to the DNLSEs
\eqref{GPspinor}. In order to find the low energy stationary
solutions, we substitute
\begin{equation}\label{spinorwave}
\begin{pmatrix}
  c^{(+)}_\eta \\
  c^{(0)}_\eta  \\
  c^{(-)}_\eta  \\
\end{pmatrix}=
   \begin{pmatrix}
  \zeta_+ \\
  \zeta_0\\
  \zeta_- \\
\end{pmatrix} \sqrt{n}\exp{[i(k a
\eta-\mu t)]}\,.
\end{equation}
Here $\mu$ is the chemical potential and
$n=\sum_{\alpha=+,0,-}|c^{(\alpha)}_\eta|^2$ is the total condensate
density that is assumed to be constant along the lattice.  The
spinor wave function,
$\vec{\zeta}^\dagger=(\zeta_+^*,\zeta_0^*,\zeta_-^*)$, satisfies the
normalization condition $\vec{\zeta}^\dagger\cdot \vec{\zeta}=1$. We
concentrate on solutions for which $\vec{\zeta}$ is constant along
the lattice.

We substitute the same Bogoliubov expansion as in \eq{BogAnsatz} for
the linearized fluctuations around the carrier wave solution
\eqref{spinorwave} in the DNLSEs \eqref{GPspinor}. This yields a $6
\times 6$ matrix $\sigma\cal{M}$, analogous to \eq{BogDeGennes2} (in
this case with $\sigma$ having three $\sigma_z$ Pauli matrices in
the diagonal), governing the dynamical stability of the system. The
matrix $\cal{M}$ is given explicitly by \eq{Mspinor} in
Appendix~\ref{app}.

As in the two-component BEC case, negative eigenvalues of the matrix
$\cal{M}$ indicate the regions of energetic instability, while the
eigenvalues of $\sigma \cal{M}$ yield the normal mode frequencies.
The imaginary parts of these normal mode frequencies represent the
strength of dynamical instabilities.

\subsection{Stability in the polar case}

\subsubsection{Dynamical stability}

In the polar case (that is energetically favored for $U_2>0$), we
consider uniform spin profiles with the average spin value zero, and
assume no Zeeman shifts for the time being $\delta_+=\delta_-=0$.
All the degenerate, physically distinguishable, ground state
solutions for $U_2>0$ may then be determined by means of the
macroscopic BEC phase $\varphi$ and a real unit vector $\dv$
defining the quantization axis of the spin. The spinor wavefunction
reads:
\begin{equation}\label{polarSoln}
\begin{pmatrix}
  \zeta_+\\
  \zeta_0  \\
  \zeta_- \\
\end{pmatrix} ={e^{i\varphi }\over \sqrt{2}}
\begin{pmatrix}
-d_{x}+id_{y} \\
\sqrt{2} d_{z} \\
d_{x}+id_{y}
\end{pmatrix} \,.
\end{equation}
As in the similar polar phase of superfluid $^{3}$He-A \cite{VOL90},
the states $(\mathbf{d},\varphi )$ and $(-\mathbf{d},\varphi +\pi )$
are identical. This can be conveniently taken into account by
considering the $\mathbf{d}$ field to define unoriented axes rather
than vectors.

The solution \eqref{spinorwave} with \eq{polarSoln} has a chemical
potential value $\mu=- 2 J \cos(k a) + U_0 n$ for any choice of
$(\mathbf{d},\varphi )$. Since also the excitations are the same for
any values of $(\mathbf{d},\varphi )$, we may choose the simplest
form of the matrix $\cal{M}$ in \eq{Mspinor}, that is obtained by
choosing $\mathbf{d}$ to point along the $z$ axis and $\varphi=0$.

\begin{figure}
\includegraphics[width=0.95\columnwidth]{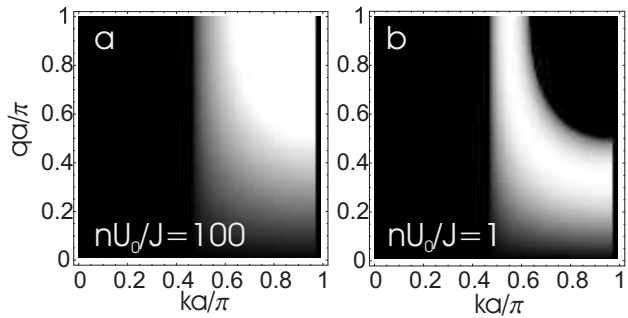}
\caption{\label{fig:polarKQ} \textbf{(a)} Largest imaginary part of
the eigenvalues (\ref{spinorPolarModeFreq}) versus $k$, holding $n
U_0=100 J$, $n U_2=4 J$. Gray scale goes from 0 to $20 J$.
\textbf{(b)} A case with a
 weaker nonlinearity $n U_0=1 J$, $n U_2=0.04 J$ Gray scale goes from 0 to $J$.}
\end{figure}

By calculating the eigenvalues of $\sigma \cal{M}$ with this
particular choice of the BEC wavefunction we obtain analytic
expressions for the normal mode energies
\begin{align}
\label{spinorPolarModeFreq}  \omega_{1 \pm}(q) & =  C_{q,k}
 \pm \sqrt{\epsilon_q \cos(k a)\big[\epsilon_q\cos(k
a) +
2 n U_0\big]} \nonumber \\
\omega_{2 \pm}(q) & =  C_{q,k}   \pm \sqrt{\epsilon_q \cos (k a)
\big[\epsilon_q \cos (k a) + 2 n U_2\big]}
\end{align}
where $\omega_{2\pm}$ are each doubly degenerate. The physical
solutions correspond to the `$+$' sign in the front of the square
root. Here again $\epsilon_q$ denotes the spectrum of an ideal,
non-moving BEC
\beq
\epsilon_{q}=4J\sin^2\({q a\over2}\)\,,
\eeq
and the Doppler shift term in the energy is given by
\beq
C_{q,k}= 2
J \sin(q a) \mathrm{sin} (k a)\,.
\eeq
It is clear from Eq.~(\ref{spinorPolarModeFreq}) that, for $U_2<U_0$
and $U_0>0$, $\omega_{1+}(q)$ drives the instability. The
$\omega_{1+}(q)$ modes are unstable when the expression inside the
square root is negative, which happens for $q$ values that satisfy:
\beq \label{unstablePolarModes} \sin^2{\(qa\over2\)}< -{n
U_0 \over 2 J \cos{(k a)}}\,.
\eeq
At least some modes are unstable whenever $k a > \pi/2$ and all the
$q$ modes are unstable when $-nU_0/(2J)<\cos{(ka)}<0$.
Figure~\ref{fig:polarKQ}(a) plots the largest imaginary part of the
mode frequencies Eq.~(\ref{spinorPolarModeFreq}) versus $q$ and $k$
for the case $n U_0/J=100$ and $n U_2/J=4$ (corresponding to
$^{23}$Na). As in the two-component case, one sees the system is
stable for $k a<\pi/2$, while for $k a>\pi/2$ one has an instability
with a growth rate linear in $q$ in the long wavelength limit before
saturating at a maximum value $\approx \sqrt{-8 n U_0 J \cos{(ka)}}
\approx 27 J$. Figure~\ref{fig:polarKQ}(b) plots a case with a much
smaller nonlinearity ($n U_0/J=1$ and $n U_2/J=0.04$). While $k
a<\pi/2$ is still the condition for stability of all the modes, one
sees that for higher values of $k$ there exists only a band of low
unstable modes in the lower $q$ region. This is due to the fact that
the RHS of (\ref{unstablePolarModes}) becomes less than unity and
thus can be exceeded by the LHS for large $q$.

\begin{figure}
\includegraphics[width=0.95\columnwidth]{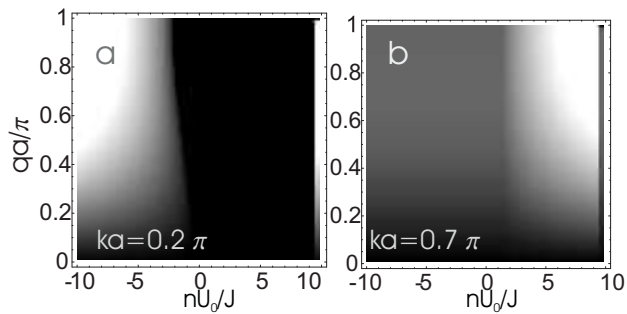}
\caption{\label{fig:polarUQ} Imaginary part of the eigenvalues
versus $n U_0$, for $n U_2=2 J$ for the condensate velocities
indicated.
 Gray scale goes from 0 to $5 J$.  }
\end{figure}

Just as we saw in the two-component case (see
Fig.~\ref{fig:twocompUQ}), these conditions are somewhat reversed
for attractive interactions $U_0<0$, as then some excitations of
$\omega_{1+}(q)$ are unstable whenever $k a<\pi/2$ and all the $q$
modes are unstable when $0<\cos{(ka)}<-nU_0/(2J)$.  The dependence
on the interaction coefficient $U_0$ is shown in
Fig.~\ref{fig:polarUQ}.  In these plots, we kept $U_2 n/J=2$
constant.   For $k a = 0.2 \pi$ (Fig.~\ref{fig:polarUQ}(a)) unstable
modes occur for negative $U_0$, while for $k a=0.7 \pi$
(Fig.~\ref{fig:polarUQ}(b)), this instability occurs for repulsive
interactions $U_0>0$.  Also, as in the two-component case, there is
an additional, weaker instability in the attractive case $U_0<0$
with $k a> \pi/2$.  This instability is driven by $\omega_{2+}(q)$
and has the same condition for instability
(\ref{unstablePolarModes}) with $U_0$ replaced by $U_2$.  The
magnitude generally reaches $\approx \sqrt{8 n U_2 J}$ for $n U_2
\gg J$.  In the case plotted in Fig.~\ref{fig:polarUQ}(b), it has a
maximum magnitude $\approx 2 J$.

The equivalence of the mode energy dependence on $U_2$ and $U_0$ in
Eq.~(\ref{spinorPolarModeFreq}) implies that whenever $U_0$ and
$U_2$ are of opposite sign and much larger than $J$, at least one of
the eigenvalues will be imaginary at any $k$.  In the previous
paragraph we discussed how this resulted in an instability for
attractive condensates $U_0<0$ with polar spin-dependent scattering
lengths $U_2>0$. Another implication of this is that polar-like
condensate solutions (\ref{polarSoln}) with $U_0>0$ are unstable for
ferromagnetic spin-dependent scattering lengths $U_2<0$. Only when
both are positive or both are negative is there a region of $k$ with
dynamical stability.

\subsubsection{Energetic stability}

\begin{figure}
\includegraphics[width=0.95\columnwidth]{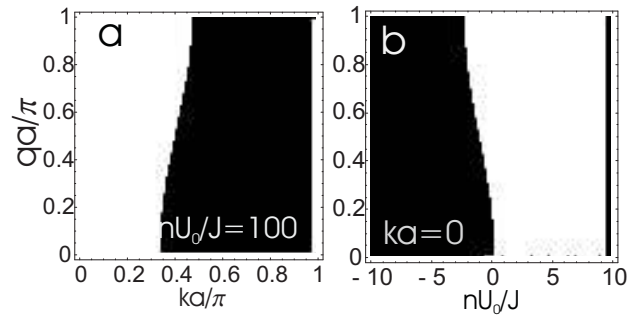}
\caption{\label{fig:polarKQenergy} \textbf{(a)} Energetic stability
(white) and instability (black) regions for $n U_0=100 J$, $n U_2=4
J$. \textbf{(b)}  Dependence of the energetic stability of
interaction coefficient $n U_0$, keeping $n U_2=2 J$. The behavior
remains qualitatively the same for non-zero $ka<\pi/2$, but with the
instability region extending into the $U_0>0$ region. }
\end{figure}

Turning now to the energetic instabilities, we also obtain analytic
results for the eigevalues of $\cal{M}$ and look for negative
eigenvalues.  We find
\begin{align}
\label{spinorPolarEnergies} \epsilon_{1 \pm}(q) & =  \epsilon_q
\cos(k a) + n U_0 \pm \sqrt{C_{q,k}^2 + n^2 U_0^2 } \nonumber \\
\epsilon_{2 \pm}(q) & =  \epsilon_q \cos(k a) + n U_2 \pm
\sqrt{C_{q,k}^2 + n^2 U_2^2 }.
\end{align}
For $U_2<U_0$ $\epsilon_{2-}$ drives the instability.
Figure~\ref{fig:polarKQenergy} plots the regions of energetic
instability. There exist unstable modes for $\cos(k a)/\sin^2(k a) <
2 J/n U_2$. For small velocities, $k a\ll 1$, this condition is
approximately equal to $\hbar k/m^*> \sqrt{nU_2/m^*}$, where
$m^*=\hbar^2/(2J a^2)$ is the effective mass of a noninteracting
BEC. This energetic instability threshold demonstrates the Landau
criterion that the velocity becomes larger than the speed of sound
(of spin waves) $a\sqrt{2JnU_2}/\hbar=\sqrt{nU_2/m^*}$.  For
$U_2>U_0$ the instability driven by $\epsilon_{1-}$ and the
condition is the same, but replacing $U_2 \rightarrow U_0$.  When
$U_0$ becomes negative, there is an additional instability, as shown
in Fig.~\ref{fig:polarKQenergy}(b).

\subsection{Stability in the ferromagnetic case}

\subsubsection{Dynamical stability}

In the ferromagnetic case (that is energetically favored for
$U_2<0$) we consider uniform spin profiles for which the magnitude
of the spin is maximized, $| \langle  {\bf F} \rangle | = 1$, which
minimizes the mean field energy (\ref{hamiltonian}). We again assume
no magnetic Zeeman shifts $\delta_+=\delta_-=0$.  As in analogous
states for superfluid liquid helium-3 \cite{VOL90}, the rotations of
the spinor axes can be used to couple physically distinguishable
ground states. Here all the degenerate states are related by spatial
rotations of the atomic spin axes and we may parametrize the spin
wavefunction as
\begin{equation} \label{ferroSoln}
\begin{pmatrix}
\zeta_+ \\ \zeta_0 \\ \zeta_{-}
\end{pmatrix}
= e^{i\phi}
\begin{pmatrix}
e^{-i\alpha}\cos^2{(\beta/2)} \\
\sin{(\beta)}/\sqrt{2} \\
 e^{i\alpha}\sin^2{(\beta/2)}
\end{pmatrix}\,,
\end{equation}
where $\alpha,\beta,\phi$ are the Euler angles. The solution
\eqref{spinorwave} with \eq{ferroSoln} has a chemical potential
$\mu=- 2 J \cos(ka) + (U_0+U_2)n$ for any chosen ground state in
\eq{ferroSoln}. The simplest form of the Bogoliubov-de Gennes matrix
$\sigma\cal{M}$ may be obtained by choosing $\beta=\alpha=\phi=0$
and substituting \eq{ferroSoln} into \eq{Mspinor} in Appendix
\ref{app}.

The mode energies are found to be:
\begin{align}
\label{spinorFerroModeFreq}  \omega_{1 \pm}(q) & =  C_{q,k}
 \pm \epsilon_q \cos(k a)  \nonumber \\
\omega_{2 \pm}(q) & = C_{q,k}  \pm \epsilon_q \mathrm{cos} (k a) \mp
2
n U_2 \nonumber \\
\omega_{3 \pm}(q) & = C_{q,k} \nonumber \\ & \pm  \sqrt{\epsilon_q
\cos(k a) \big[\epsilon_q\cos(k a) + 2 n(U_0+U_2) \big]}
\end{align}
The dynamical instabilities are driven entirely by $\omega_{3+}$ and
the only difference with the polar case
Eq.~(\ref{spinorPolarModeFreq}) is the replacement $U_0$ and $U_2$
individually by the sum $U_0+U_2$.   This dependence can be
understood from the fact that with $| \langle  {\bf F} \rangle | =
1$ the total nonlinearity is $\propto U_0+U_2$ and so this quantity
determines the attractive or repulsive character of the condensate.
Thus for $U_0>0$ and $|U_2| \ll |U_0|$,  the instability diagrams
qualitatively similar to Fig.~\ref{fig:polarKQ}.

\begin{figure}
\includegraphics[width=0.95\columnwidth]{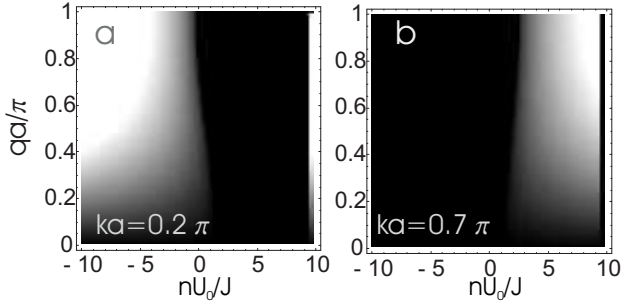}
\caption{\label{fig:ferroUQ} Imaginary part of the eigenvalues
(\ref{spinorPolarModeFreq}) versus $n U_0$, holding $n U_2=-2 J$,
for the two condensate velocities indicated. Gray scale goes from 0
to $5 J$. }
\end{figure}

Differences between the ferromagnetic and polar cases become clear
when one examines the instability dependence on $U_0$. This is shown
for the ferromagnetic case in Fig.~\ref{fig:ferroUQ}, where we keep
$n U_2 = -2 J$ constant, and should be contrasted with the polar
case, Fig.~\ref{fig:polarUQ}. For $k a < \pi/2$ (the figure shows $k
a=0.2 \pi$), an instability occurs for $U_0 < -U_2$ and increases
with greater $|U_0|$ while for $k a > \pi/2$ (the figure shows $k
a=0.7 \pi$), the instability occurs for $U_0>-U_2$. An important
difference from the polar case is that, for $k a>\pi/2$, there is no
instability for $U_0<0$. In addition, the instability border occurs
at $U_0=-U_2$ rather than at $U_0=0$. Finally, from
Eq.~(\ref{spinorFerroModeFreq}), we note that a ferromagnetic BEC
solution (\ref{ferroSoln}) with $U_0>0$ and a {\it polar}
spin-dependent scattering length $U_2>0$ can be dynamically stable,
in contrast to a polar solution with ferromagnetic scattering
length, as discussed above.

\subsubsection{Energetic stability}

\begin{figure}
\includegraphics[width=0.95\columnwidth]{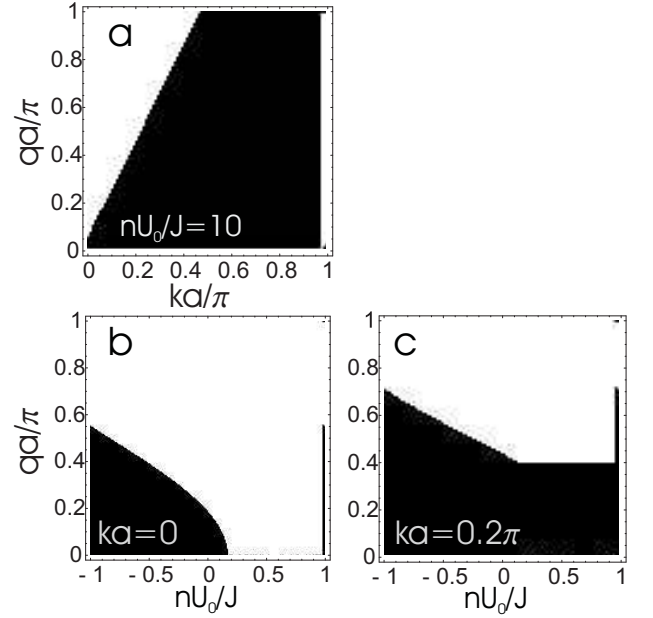}
\caption{\label{fig:ferroKQenergy} \textbf{(a)} Energetic stability
regions for the ferromagnetic case with $^{87}$Rb parameters
$U_0=100 J$, $U_2=-0.36 J$. \textbf{(b)-(c)} Energetic stability
versus the interaction coefficient $U_0$, keeping $U_2=-2 J$.}
\end{figure}

The energy eigenvalues in the ferromagnetic case are
\begin{eqnarray}
\label{spinorFerroEnergies} \epsilon_{1 \pm}(q) & = & \epsilon_q
\cos(k a)
\pm C_{q,k}  \nonumber \\
\epsilon_{2 \pm}(q) & = & \epsilon_q
\cos(k a) -2 n U_2 \pm C_{q,k}  \nonumber \\
\epsilon_{3 \pm}(q) & = & \epsilon_q \cos(k a) + n (U_0+U_2)
\nonumber
\\ & & \pm \sqrt{C_{q,k}^2 + n^2 (U_0+U_2)^2  }.
\end{eqnarray}
\noindent Unlike the polar case, there is one energy eigenvalue
$\epsilon_{1-}$ corresponding to a pure (Doppler-shifted) kinetic
energy.  This gives rise to energetic instabilities for $\sin^2(q
a/2)/\sin(qa) < \tan(ka)$, as plotted in
Fig.~\ref{fig:ferroKQenergy}(a).  Though no dynamic instability
exists except for much higher $k$, ferromagnetic spinor BECs are
subject to this energetic instability in the presence of thermal
excitation for any non-zero $k$.  We note that for $k a>0.5 \pi$ all
$q$ modes are unstable.

Finally, for attractive condensates ($U_0+U_2<0$) there are
additional regions of energetic instability from $\epsilon_{3-}$.
Fig.~\ref{fig:ferroKQenergy}(b-c) shows the dependence versus $U_0$.
For $k=0$ there are unstable modes for all attractive condensate
scattering length cases.  A moving condensate ($k a=0.2 \pi$ is
shown in Fig.~\ref{fig:ferroKQenergy}(c)), increases the region of
unstable $q$ modes for attractive condensates.  The band of
energetic instability at low $q$ for $U_0>0$ in
Fig.~\ref{fig:ferroKQenergy}(c) is simply the Doppler induced
energetic instability discussed above.

\subsection{Effects of Zeeman splitting}

When Zeeman splittings due an external magnetic field [$\delta_\pm$
in Eq.~(\ref{GPspinor})] are non-zero, the symmetry of the polar and
ferromagnetic solutions, Eqs.~\eqref{polarSoln}
and~\eqref{ferroSoln}, breaks down and we find a new set of
steady-state solutions.  Here we examine these solutions and again
calculate the dynamic stability of these various solutions,
particularly noting how the stability varies with the {\it linear}
and {\it quadratic} Zeeman shifts, which we denote, respectively, as
$\tilde\delta=(\delta_+-\delta_-)/2$ and
$\bar\delta=(\delta_++\delta_-)/2$.  The linear Zeeman shifts are
$\tilde\delta=2 g_F \mu_B B=2 \big((2 \pi) 1.4~\mathrm{MHz/G}
\big)B$, where $g_F$ is the Land$\acute{e}$ factor, and is $-1/2$
for the ground-state $F=1$ manifold of $^{87}$Rb and $^{23}$Na.  For
Zeeman shifts substantially smaller than the hyperfine splitting,
which is our interest here, the quadratic shifts $\bar{\delta}$ are
typically smaller than the linear shifts and can be extracted from
the Breit-Rabi formula \cite{COR77}. For alkali atoms the quadratic
shift is positive, but it can generally be of either sign. The level
shifts in a spin-1 BEC may also be engineered in other ways, e.g.,
by using off-resonant microwave fields that generate
electromagnetically-induced level splittings \cite{GER06}, allowing
essentially arbitrary experimentally prepared level shifts for
$\tilde\delta$ and $\bar\delta$. Note that the linear Zeeman shift
does not affect the energy conservation of a spin-changing collision
$|m_F=0,m_F=0 \rangle \leftrightarrow |m_F=+1,m_F=-1 \rangle$, while
the quadratic Zeeman shift does, and thus plays an important role in
the stability properties.

One of the steady-state solutions in the presence of the Zeeman
splitting has the chemical potential $\mu=\mu_0=-2 J\cos{(k a) } +n
U_0$ and reads:
\begin{equation} \label{polar1}
\vec\zeta = e^{i\varphi}
\begin{pmatrix}
0 \\
1 \\
0
\end{pmatrix}\,.
\end{equation}
This solution forms a subset of the polar solutions
(\ref{polarSoln}) in the absence of the magnetic splitting, with the
$\dv$ pointing along the $z$ direction.

The stability is again analyzed by substituting the steady-state
solution [\eq{polar1}] into \eq{Mspinor}. By calculating the
eigenvalues of $\sigma \cal{M}$ we obtain analytic expressions for
the normal mode energies, as in \eq{spinorPolarModeFreq}. Here
$\omega_{1 \pm}(q)$ remains unchanged in the presence of the Zeeman
splitting and
\begin{align}
\label{spinorPolarModeFreq2}
%&\omega_{1 \pm}(q) =  C_{q,k}
%\pm \sqrt{\epsilon_q \cos(k a)\big[\epsilon_q\cos(k a) +
%2 U_0 n\big]} \nonumber \\
& \omega_{2 \pm}(q)  =  C_{q,k} \pm \tilde\delta \pm P_{q,k}\nonumber \\
& \omega_{3 \pm}(q) =   C_{q,k} \mp \tilde\delta \pm P_{q,k}\nonumber \\
& P_{q,k} = \sqrt{\big[\epsilon_q \cos (k a)+\bar\delta\big]
\big[\epsilon_q \cos (k a) + 2 n U_2+\bar\delta\big]}\,,
\end{align}
The linear splitting lifts the degeneracy between $\omega_{2
\pm}(q)$ and $\omega_{3 \pm}(q)$ in \eq{spinorPolarModeFreq} and the
quadratic splitting introduces an energy gap $\bar\delta$ in the
single-particle phonon mode spectrum $\epsilon_q \cos (k a)$ in
$\omega_{2 \pm}(q)$ and $\omega_{3 \pm}(q)$.

The dynamical stability will then be governed by the sign of the
square root argument of $P_{q,k}$ and is seen to be unaffected by
the linear Zeeman shift. The condition for the existence of unstable
modes is $-\epsilon_q \cos (k a)-2 n U_2<\bar\delta< -\epsilon_q
\cos (k a)$ for $U_2>0$, and $-\epsilon_q \cos (k a)<\bar\delta<
-\epsilon_q \cos (k a)-2 n U_2$ for $U_2<0$. If $U_2>0$ and $k
a<\pi/2$, Eq.~(\ref{spinorPolarModeFreq2}) predicts dynamical
stability for positive quadratic shift $\bar{\delta}>0$, and
instability at some $q$ for $\bar{\delta}<0$. As in previous cases,
we find much larger instabilities when $ka>\pi/2$ (for $U_0>0$) from
$\omega_{1 \pm}$.

Similarly, we obtain the eigenvalues of ${\cal M}$ as in
\eq{spinorPolarEnergies}. Here $\epsilon_{1 \pm}(q)$ is unchanged
and
\begin{align}
\label{spinorPolarEnergies2} \epsilon_{2 \pm}(q) & =  \epsilon_q
\cos(k a) + n U_2 +\bar\delta \pm \sqrt{(C_{q,k}+\tilde\delta)^2 +
n^2 U_2^2 }\nonumber\\ \epsilon_{3 \pm}(q) & =  \epsilon_q \cos(k a)
+ n U_2 +\bar\delta \pm \sqrt{(C_{q,k}-\tilde\delta)^2 + n^2 U_2^2
}\,.
\end{align}
\begin{figure}
\includegraphics[width=0.95\columnwidth]{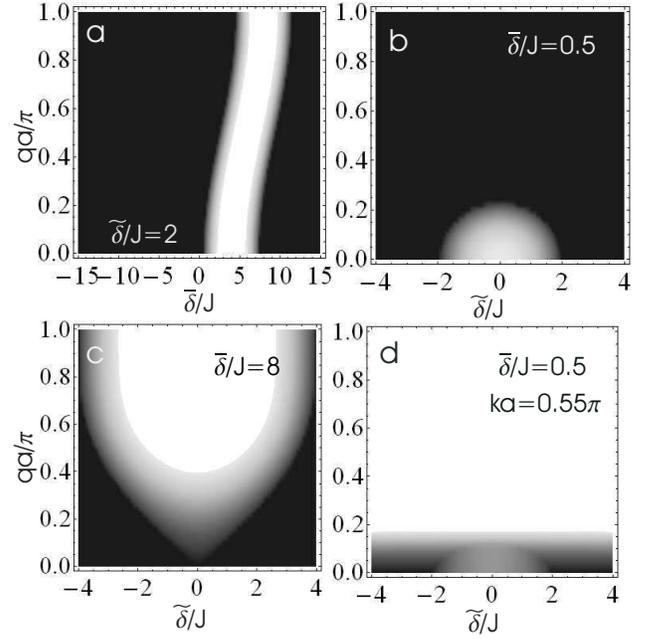}
\caption{\label{fig:polarZeeman} Largest imaginary parts of the
normal mode energies for the polar-like solution Eq.~(\ref{polar2}).
In all plots, $U_0=100 J$ and $U_2=0.04 U_0$ (corresponding to
$^{23}$Na). Plots are on a gray scale of 0 to $3 J$.  \textbf{(a)}
Dynamic instability versus $\bar{\delta}$ for linear Zeeman shift
$\tilde\delta=2J$ and $k=0$. \textbf{(b-c)} Versus linear shift
$\tilde{\delta}$ for quadratic Zeeman shifts $\bar{\delta}=0.5 J$
and  $8 J$, with $k=0$. \textbf{(d)} Versus $\tilde{\delta}$ with $k
a=0.55 \pi$.}
\end{figure}

We find another steady-state solution to the DNLSEs \eqref{GPspinor}
with the chemical potential $\mu=\mu_0+\bar\delta$ that reads
\begin{equation} \label{polar2}
\vec\zeta = {e^{i\varphi}\over\sqrt{2}}
\begin{pmatrix}
e^{-i\gamma}\sqrt{1-{\tilde\delta/ nU_2}} \\
0 \\
e^{i\gamma}\sqrt{1+{\tilde\delta/ nU_2}}
\end{pmatrix} \,.
\end{equation}
This solution only exists for sufficiently small linear Zeeman
shifts $|\tilde{\delta}|<|n U_2|$. For $\tilde\delta=0$, \eq{polar2}
coincides with the subset of solutions to \eq{polarSoln}. Although
for small $\tilde\delta\neq0$ the solution \eqref{polar2} is still
close to the polar state of \eq{polarSoln}, the spin expectation
value is no longer zero, $\<{\bf F}\>= -\tilde\delta/(nU_2) {\bf
\hat{z}}$, with $\dv$ restricted on the $xy$ plane and the induced
$\<{\bf F}\>$ pointing along the $z$ axis~\cite{comment2}. At the
boundary of the validity of \eq{polarSoln}, $\tilde{\delta}=\pm n
U_2$, we obtain $\<{\bf F}\>= \mp {\bf \hat{z}}$, as in a
ferromagnetic state, so that \eq{polar2} in fact interpolates
between the polar and the ferromagnetic solutions. Moreover, we
again find analytic solutions for the normal mode energies:
\begin{widetext}
\begin{align}
\label{spinorPolarModeFreqZeeman}  \omega_{1 \pm}(q) & =  C_{q,k}
 \pm \sqrt{\epsilon_q \cos(k a)\big[\epsilon_q\cos(k
a) +
2 n U_2-2\bar\delta\big]+\tilde\delta^2+\bar\delta^2 - 2nU_2\bar\delta} \nonumber \\
\omega_{2 \pm}(q) & =  C_{q,k}\pm \sqrt{\epsilon_q
\cos (k a) \big[\epsilon_q \cos (k a) + n(U_0+U_2)\big]+L_{q,k}}\nonumber \\
\omega_{3 \pm}(q) & =  C_{q,k} \pm \sqrt{\epsilon_q
\cos (k a) \big[\epsilon_q \cos (k a) + n(U_0+U_2)\big]-L_{q,k}}\nonumber \\
L_{q,k} &= \sqrt{\epsilon_q^2 \cos^2 (k a)\big[ n^2(U_0-U_2)^2+4
\tilde\delta^2 U_0/ U_2\big]}\,.
\end{align}
\end{widetext}

For $U_0,U_2>0$ and $k a <\pi/2$, whenever the solution
\eqref{polar2} exists (i.e., when $|n U_2|\geq |\tilde\delta|$), the
dynamical instabilities are solely driven by $\omega_{1 \pm}(q)$.
Moreover, under these conditions the mode $\omega_{1 \pm}(q)$ is
dynamically stable when \eq{polar2} is energetically favorable to
\eq{polar1} (i.e., when $\bar\delta<0$). In
Fig.~\ref{fig:polarZeeman} we show some stability diagrams for the
parameters of $^{23}$Na.  The instability dependence on the
quadratic Zeeman shift for a particular linear Zeeman shift is shown
in Fig.~\ref{fig:polarZeeman}(a). This diagram may be understood by
noting that the mode $\omega_{1 \pm}(q)$ exhibits a nonvanishing
imaginary part, if
\begin{align}
& c_-<\bar\delta < c_+,\nonumber\\ & c_\pm\equiv \epsilon_q \cos(k
a)+nU_2 \pm \sqrt{n^2 U_2^2-\tilde\delta^2}\,.\label{zeemaninst}
\end{align}
This forms an instability stripe with a width $2(n^2
U_2^2-\tilde\delta^2)^{1/2}$ in Fig.~\ref{fig:polarZeeman}(a) and
for larger values of $\bar{\delta}$ the system stabilizes again. We
observe the stripe position shift in $\bar{\delta}$ by $4J$ from the
$q a=0$ to $q a = \pi$. We also calculated the energetic stability
of \eq{polar2} and found that in Fig.~\ref{fig:polarZeeman}(a) the
region to the left of the stripe is energetically stable, while the
entire region to the right of the stripe (i.e., large
$\bar{\delta}$) is energetically unstable.

While not obvious in Fig.~\ref{fig:polarZeeman}(a) there is a small
region of stability for positive $\bar\delta$.  This is seen more
easily by plotting the instability versus the linear Zeeman shift.
In Figs.~\ref{fig:polarZeeman}(b-d) we plot this for the entire
range of validity of \eq{polar2} ($|\tilde\delta|\leq |n U_2|$) for
various $\bar\delta$. For small quadratic Zeeman shift, there exists
a range of linear Zeeman shifts which stabilize the system, as it
approaches the ferromagnetic state. For larger quadratic shifts this
range shrinks, until eventually the system is unstable at all
possible $\tilde{\delta}$, as in Fig.~\ref{fig:polarZeeman}(b).

The effect of larger $k$ is generally simply to stretch the region
of instability to larger ranges of $q$ for each case, as the effect
of the $\epsilon_q$ shift in \eq{zeemaninst} vanishes. For $ka>0.5
\pi$ (with $U_0>0$) the usual, and much larger, instability for
large $k$, seen in previous cases, dominates the stability diagram.
Such a case is plotted in Fig.~\ref{fig:polarZeeman}(d).

In the absence of the Zeeman splitting the polar state
\eqref{polarSoln} is always dynamically unstable for $U_2<0$. For
the state \eqref{polar2}, even close to the polar state, this is no
longer the case. For $U_0>0$ and $U_2<0$, \eq{zeemaninst} represents
the entire unstable region, provided that $4\tilde\delta^2
|U_0/U_2|\leq n^2(U_0-U_2)^2$ and $|U_0|>|U_2|$.

In the presence of the Zeeman splitting the ferromagnetic state
(\ref{ferroSoln}) is modified to
\begin{equation} \label{ferro1}
\vec\zeta = e^{i\phi}
\begin{pmatrix}
1 \\
0 \\
0
\end{pmatrix}\,,
\end{equation}
with the chemical potential $\mu=\mu_0+nU_2+ \delta_+$, or to an
analogous ferromagnetic state with $(\zeta_+,\delta_+)$ interchanged
with $(\zeta_-,\delta_-)$. Although \eq{ferro1} is a steady-state
solution to the DNLSEs \eqref{GPspinor} for any values of
$\tilde\delta$, we also find that the solution \eqref{polar2} has
the limit \eq{ferro1} at the boundary of the validity region
$\tilde\delta=-nU_2$. At the other limit of the validity of
\eq{polar2} (at $\tilde\delta=n U_2$) we recover the other
ferromagnetic state, defined by $(\zeta_-,\delta_-)$. Moreover, the
solution \eqref{ferro1} is energetically favorable to \eq{polar1}
when $nU_2/2+ \delta_+<0$ and to \eq{polar2} when $nU_2/2+
\tilde\delta<0$.

We find that the normal mode energies and the eigenvalues of ${\cal
M}$ corresponding to \eq{ferro1} are obtained from the non-Zeeman
shifted mode frequencies, Eqs.~\eqref{spinorFerroModeFreq}
and~\eqref{spinorFerroEnergies}, by shifting the single-particle
excitation energies: $\epsilon_q \cos (k a)\rightarrow \epsilon_q
\cos (k a)-\delta_+$ in $\omega_{1 \pm}(q)$ and $\epsilon_{1
\pm}(q)$; and $\epsilon_q \cos (k a)\rightarrow \epsilon_q \cos (k
a)-2\tilde\delta$ in $\omega_{2 \pm}(q)$ and $\epsilon_{2 \pm}(q)$.
The energies $\omega_{3 \pm}(q)$ and $\epsilon_{3 \pm}(q)$ are
unchanged. Because it is $\omega_{3 \pm}$ which drives the dynamical
instability, the stability diagram is unchanged from that of
\eq{spinorFerroModeFreq}.
\begin{figure}
\includegraphics[width=0.95\columnwidth]{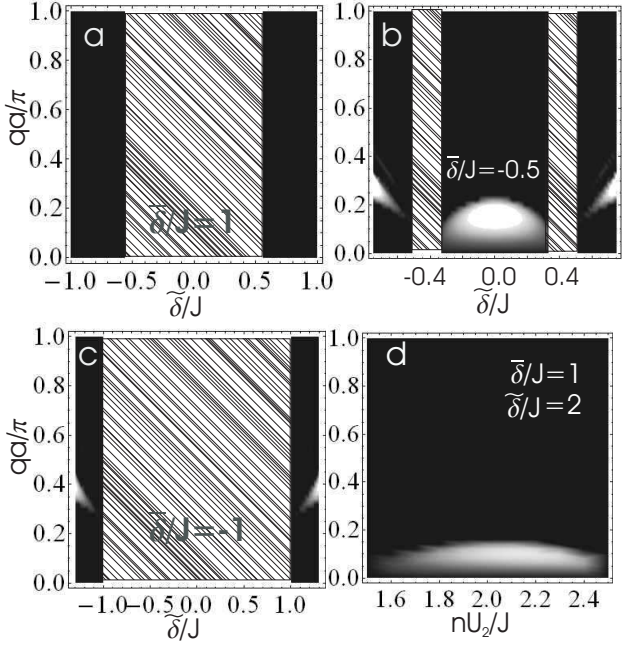}
\caption{\label{fig:ferroZeeman} Largest imaginary parts of the
normal mode energies for the solution Eq.~(\ref{ferro2}). In all
plots, $n U_0=100 J$.   Hatched areas represent regions where the
condensate solution is not valid. \textbf{(a)-(c)} Dynamic
instability versus linear Zeeman shift and $q$ for $U_2=-0.0036 U_0$
(corresponding to $^{87}$Rb). Gray scale runs from (a) 0 to $0.3 J$,
(b) 0 to $0.3 J$, (c) 0 to $0.2 J$. In (a) we emphasize that there
is a range of $\tilde{\delta}$, on each side of the instability
region, stable for all $q$. \textbf{(d)} Dynamic instability versus
spin-dependent scattering length $U_2$ for $\tilde{\delta}=2 J$ and
$\bar{\delta}=J$. Gray scale from 0 to $0.2 J$. }
\end{figure}

We find an additional ferromagnetic-like steady-state solution to
the DNLSEs \eqref{GPspinor} with $\mu=\mu_0+nU_2+
(\bar{\delta}^2-\tilde{\delta}^2)/2 \bar{\delta}$, that reads
\begin{align} \label{ferro2} \zeta_\pm &= e^{i\gamma_\pm}\delta_\mp
\sqrt{{2 nU_2 \bar{\delta}+ \bar{\delta}^2-\tilde{\delta}^2\over 8
nU_2
\bar{\delta}^3}},\nonumber\\
\zeta_0 & =e^{i(\gamma_++\gamma_-)/2}
\sqrt{1-|\zeta_+|^2-|\zeta_-|^2}\,.
\end{align}
This solution only exists if the expressions inside the square roots
of \eq{ferro2} are positive, i.e.,
\beq 0\leq {2 nU_2 \bar{\delta}+
\bar{\delta}^2-\tilde{\delta}^2\over 8 nU_2 \bar{\delta}^3} \leq
{1\over 2 (\tilde\delta^2+\bar\delta^2)}\,.\label{limitf}
\eeq
This condition gives rise to a ranges of validity for $U_2$ in terms
of two parameters
$\beta_-=(\tilde{\delta}^2-\bar{\delta}^2)/2\bar{\delta}$ and
$\beta_+=(\tilde{\delta}^2+\bar{\delta}^2)/2\bar{\delta}$.   In the
most common case that the linear Zeeman shift is larger in magnitude
($|\tilde{\delta}|>|\bar{\delta}|$) these inequalities give a finite
range: $\beta_- < n U_2 < \beta_+$ for $\bar{\delta}>0$ and $\beta_+
< n U_2 < \beta_-$ for $\bar{\delta}<0$.

If $|\tilde{\delta}|<|\bar{\delta}|$, the inequalities change
direction, giving instead an intermediate range of $U_2$ where
(\ref{ferro2}) is {\it not} valid.  In particular, the requirement
for validity for $\bar{\delta}>0$ is $n U_2<\beta_-$ or $n
U_2>\beta_+$, while for $\bar{\delta}<0$ it is $n U_2<\beta_+$ or $n
U_2>\beta_-$.

At the lower limit of \eq{limitf} ($nU_2=\beta_-$) the solution
\eqref{ferro2} coincides with the polar solution \eqref{polar1} with
$\<{\bf F}\>=0$, while at upper limit ($nU_2=\beta_+$) it equals
\eq{polar2}. In general, the spin for the solution (\ref{ferro2}) is
non-vanishing
\beq
|\<{\bf
F}\>|^2={2\bar\delta^2(\tilde\delta^2+2
n^2U_2^2)-\bar\delta^4-\tilde\delta^4\over 4\bar\delta^2n^2U_2^2}\,.
\eeq
This condensate solution (\ref{ferro2}) is energetically favorable
to (\ref{ferro1}) for {\it positive} quadratic shifts
$\bar{\delta}>0$, the opposite relationship of solution
(\ref{polar2}) to (\ref{polar1}).

We computed the dynamical and energetic stability of this solution.
Figures~\ref{fig:ferroZeeman}(a)-(c) show the dynamical instability
strengths as a function linear Zeeman shift for several quadratic
shifts and Rb-87 scattering length ($U_2=-0.0036 U_0$).  The hatched
areas and all $|\tilde{\delta}|$ larger than the range of these
plots are regions where the condensate solution (\ref{ferro2}) is
not valid. For $\bar{\delta}>0$ (Fig.~\ref{fig:ferroZeeman}(a)) we
see the valid regions where the solution exists are dynamically
stable. For $\bar{\delta}<0$ (Figs.~\ref{fig:ferroZeeman}(b)-(c))
there are always some unstable modes $q$.  However note the
interesting behavior that for large $|\tilde{\delta}|$, there exist
only small bands of unstable $q$ and weak instability (note the
range of the plots in the caption). The system is {\it
energetically} unstable for $q$ in the regions overlapping and below
the small dynamical instability bands.

In Fig.~\ref{fig:ferroZeeman}(d) we plot the instability strength
versus $U_2$ over its range of validity, which in this case is $1.5
J < n U_2 < 2.5 J$.  Here we see a weak band of instability for low
$q$ in the range of $U_2$ where the solution exists.

\section{Experimental considerations} \label{experimental}

In the experimental realizations of optical lattice systems,
ultra-cold atoms have been trapped in a combined optical lattice and
a harmonic trap. The transport properties may then be studied by
suddenly displacing the harmonic trap, e.g., by using a magnetic
field gradient. This excites dipolar oscillations of atoms along the
lattice direction with the maximum velocity proportional to the
harmonic trap displacement \cite{BUR01,FER05}. The other alternative
is to use a moving-standing wave, so that the atoms are trapped
close to the harmonic trap minimum and experience a moving optical
lattice potential \cite{FAL04,SAR05}. The advantage of the latter
technique is that the velocity of the atoms with respect to the
lattice is constant. In such transport experiments the dynamical
instabilities may typically be observed on much shorter time scales
than the energetic ones and the rate of the energetic instability to
have an observable effect can be controlled by increasing the size
of the thermal atom cloud \cite{SAR05}.

A two-component ultra-cold $^{87}$Rb vapor has also been trapped in
a spin-dependent lattice using two counter-propagating laser beams
with linear polarizations \cite{MAN03}. The two species experience
different $\sigma_+$ and $\sigma_-$ polarized optical lattices where
the separation between the lattice potentials can be controlled by
changing the angle between the linear polarization vectors.

The techniques developed for investigating dynamical and energetic
instabilities in a single component case could be adapted to our
proposed two-component BEC studies. A spin-dependent lattice
potential may be used to control the value of the intraspecies
interaction strength $U_{12}$ by modifying the spatial overlap
integral between the lattice site wavefunctions of the two species.
Moreover, the various intra- and inter-species scattering lengths
and the two species Feshbach resonances \cite{KOK07} in
two-component BEC systems make them a very rich area for
experimental exploration. In addition to the two-component $^{87}$Rb
vapor \cite{MAN03}, two-component BECs have been experimentally
realized in optical lattices using a $^{41}$K--$^{87}$Rb mixture
\cite{CAT07}.

Different superfluid velocities for the two species can be realized
in such a system by moving the two lattice potentials at different
speed. The disadvantage of this scheme is that it would make the
intraspecies interaction strength $U_{12}$ time-dependent.
Displacing the harmonic trapping potentials of the two BECs
different distances, in such a way that at the end the traps are
perfectly overlapping, could be used to realize two BECs undergoing
dipolar oscillations in phase with different amplitudes. Perhaps the
easiest method to measure the reversed phase separation instability
in a two-species BEC, as discussed in Section~\ref{diffsigns} and
Appendix~\ref{appdynamical}, is to move the lattice potentials of
the both species at the same speed and to use light-stimulated
coherent Bragg diffraction \cite{KOZ99} to change the velocity of
one of the BECs. For sufficiently small velocities of the Bragg
diffracted BEC, the atomic clouds of the two BECs overlap long
enough for the dynamical instabilities to have an observable effect.

Atomic $^{87}$Rb spin-1 gases have also been loaded to optical
lattices \cite{WID05}. In such a system the linear and quadratic
Zeeman shifts could be modified, e.g., by using off-resonant
microwave field-induced level shifts \cite{GER06}. This would allow
the studies of the stability properties of different steady-state
solutions presented here.

\section{Conclusions} \label{conclusion}

We studied the transport properties of two-component and spinor
atomic BECs in optical lattices using the discrete nonlinear
Schrodinger equations, obtained in the tight-binding approximation
to the lattice system. The classical GP theory is valid in optical
lattices at low temperatures if the effective 1D nonlinearity is not
too large, the atom number not too small, or the lattice potential
not too deep \cite{ISE05}. In particular, we studied both the
dynamical and energetic stability of homogenous Bloch wave solutions
to the DNLSEs for the condensates by analyzing the linearized
perturbations around the carrier wave. In the case of the dynamical
instabilities this involved finding the eigenvalues (normal mode
energies) of the corresponding Bogoliubov-de Gennes equations [the
matrix $\sigma{\cal M}$ in \eq{BogDeGennes2}] and in the case of
energetic instabilities finding the eigenvalues of the second order
perturbations in the energy functional (the matrix ${\cal M}$). Our
steady-state Bloch wave ansatz allowed for magnetic Zeeman level
shifts and even for two different velocities of the condensates in
the two-component case. Our study discusses a large number of cases
and points out how the spin degree of freedom can affect the
stability properties.

In the two-component case we analyzed and fully characterized the
dynamical and energetic instabilities for the general case of the
two BECs exhibiting arbitrary hopping amplitudes, interaction
strengths, superfluid velocities, and spatial overlap. Simple
analytic expressions for the normal mode energies [\eq{2becf}] and
the dynamical stability criteria were obtained in the important case
of the two BECs having the same atom current (even when the
velocities may differ).

For the case that $\cos(k_1a)$ and $\cos(k_2a)$ exhibit equal sign
for the two BEC carrier wavenumbers $k_1,k_2$ (see
Sec.~\ref{samesigns}), we found that the instability diagram
contains contributions from: (1) the high velocity instability,
determined by \eq{insta}, (which is analogous to that of a
single-component BEC in a lattice), and (2) a weaker
phase-separation instability [\eq{phasesep}], which occurs when
$U_{12}^2> U_{11} U_{22}$. However, as shown in
Appendix~\ref{appdynamical} and in Sec.~\ref{diffsigns}, an
interesting case arises when one allows different condensate
velocities of the two-components, so that $\cos(k_1a)$ and
$\cos(k_2a)$ exhibit different sign (the effective masses of the two
components exhibit different signs). Firstly, the high velocity
instability conditions (which depends on the velocities, hopping
amplitudes, atom numbers, and interaction strengths of the two BECs)
indicate that the presence of the other condensate component can
stabilize the superfluid flow of an otherwise unstable condensate
(that exceeds the critical velocity of a single-component BEC).
Secondly, the phase separation stability criteria can be reversed
for particular sets of parameters and the entire dynamically {\it
stable} regime exists for $U_{12}^2> U_{11} U_{22}$; see
Appendix~\ref{appdynamical} and Figs.~\ref{fig:twocompU12Q}(c-d).

For the spin-1 BEC case, we also obtained analytic expressions for
the dynamical and energetic instabilities in several cases of
interest. In the absence of the Zeeman level shifts the normal mode
energies in the polar and ferromagnetic ground state manifolds are
simplified and the two cases differ when $U_0<0$ and for relatively
large spin-dependent interactions $|U_2|\sim|U_0|$. In particular,
the polar case tends to exhibit more regions of instability as $U_0$
and $U_2$ separately contribute [see \eq{spinorPolarModeFreq}],
while in the ferromagnetic case, it is the sum $U_0+U_2$ which is
important; see \eq{spinorFerroModeFreq}. This allows for dynamical
stability of ferromagnetic solutions for $U_0<0$, and even for polar
spin-dependent scattering lengths ($U_2>0$). Also, we found that,
unlike the polar case, the ferromagnetic solution will have some
{\it energetic} instability for any finite BEC velocity due to the
existence of a pure kinetic energy eigenvalue $\epsilon_{1 \pm}$
[Eq.~(\ref{spinorFerroEnergies})].

In the presence of the linear and quadratic Zeeman level shifts we
find a new set of steady-state Bloch wave solutions, describing the
superfluid flow in spin-1 BECs. While the polar-like solution
\eqref{polar1} and the ferromagnetic-like solution \eqref{ferro1}
form subsets of the corresponding solutions in the absence of the
Zeeman splitting, this is not the case for the steady-state
solutions \eqref{polar2} and \eqref{ferro2}. The solution
\eqref{polar2} exhibits a nonvanishing spin vector $\< {\bf F}\>$
pointing along the magnetic field and interpolates between the polar
and the ferromagnetic solutions. The solution \eqref{ferro2} only
exists in the presence of sufficiently large Zeeman shifts and in
this sense represents an entirely novel state.

We analyzed the stability conditions for all spin-1 Bloch wave
states. For condensate solutions unique to the presence of Zeeman
shift, Eqs.~\eqref{polar2} and~\eqref{ferro2}, the stability
diagrams were presented in Figs.~\ref{fig:polarZeeman}
and~\ref{fig:ferroZeeman}, respectively. For the parameters of
$^{23}$Na at low velocities we found the solution \eq{polar2} to be
dynamically stable for negative quadratic shifts. Even for positive
quadratic shift, a sufficiently large linear shift can stabilize it.
Moreover, the solution (\ref{polar2}) can be stable for
ferromagnetic scattering coefficients ($U_2<0$), even close to the
polar state. For the parameters of $^{87}$Rb the solution
\eq{ferro2} can be energetically and dynamically stable for positive
quadratic Zeeman shifts and sufficiently large linear Zeeman shifts.

The phenomena discussed in this paper should be applicable to
current and future experiments with multiple-component BECs in
optical lattices, and we discussed some of the important
considerations for the experimental realization. We concentrated on
the stability studies of moving Bloch wave solutions in the lattice.
An interesting theoretical extension of this work is to consider
inhomogenous condensate solutions, such as soliton-like structures
in spinor BECs \cite{DAB07}. One could also investigate the effect
of spin-dependent lattice potentials with spatially inhomogeneous
profiles for the hopping amplitude, for instance dimerization, along
the lattice \cite{RUO02}.

\acknowledgements One of us (JR) acknowledges discussions with L.\
De Sarlo, M.\ Inguscio, and F.\ Minardi.  ZD was supported in this
work by the Office of Naval Research.

\appendix

\begin{widetext}

\section{Bogoliubov-de Gennes matrices} \label{app}

In the two-component case, the Bogoliubov-de Gennes equation, upon
substitution of the the ansatz (\ref{BogAnsatz}) into (\ref{gpe2}),
we get (\ref{BogDeGennes2}) with:
\begin{equation} \label{MtwoComp}
\cal{M} = \left(%
 \begin{array}{cccc}
  K_{1+} + \bar{U}_1 & - \bar{U_1} &  \bar{U}_{12} & -\bar{U}_{12} \\
  -\bar{U}_1 &   K_{1-} + \bar{U}_1 & -\bar{U}_{12} & \bar{U}_{12} \\
  \bar{U}_{12} & -\bar{U}_{12} & K_{2+} + \bar{U}_{2} & -\bar{U}_2 \\
  -\bar{U}_{12} &\bar{U}_{12} & -\bar{U}_{2} & K_{2-} +\bar{U}_2 \\
\end{array}%
\right)
\end{equation}
\noindent where $K_{j\pm} = (4 J_j/\hbar) \sin^2(q a/2) \cos(k_j a)
\pm (2 J_j/\hbar) \sin(q a) \sin(k_j a)$, $\bar{U}_j = U_j n_j/
\hbar$, and $\bar{U}_{12} = U_{12} \sqrt{n_1 n_2}/ \hbar$.

In the general spinor case, we obtain the $\cal{M}$ for the spinor
wavefunction and the Bogoliubov expansion (\ref{BogAnsatz}), into
the DNLSEs (\ref{GPspinor}):
\begin{equation}\label{Mspinor}
  \cal{M} = \left(%
\begin{array}{cccccc}
   M_-^{(+)} & w_{++} & f_{0+}^{0-} & w_{0+} & r_{-+} &  g_{-+}^{0} \\
  (w_{++})^* & M_+^{(+)}  & (w_{0+})^* & (f_{0+}^{0-})^* & (g_{-+}^{0})^*  & (r_{-+})^* \\
   (f_{0+}^{0-})^* & w_{0+}  & M_-^{(0)} & h^{-+}_{0} & (f_{0-}^{0+})^*   & w_{0-}   \\
   (w_{0+})^* & f_{0+}^{0-}  & (h^{-+}_{0})^* & M_+^{(0)} &  (w_{0-})^*  & f_{0-}^{0+}  \\
  (r_{-+})^* & g_{-+}^{0} & f_{0-}^{0+} & w_{0-} &  M_-^{(-)} &  w_{--} \\
  (g_{-+}^{0})^* & r_{-+} & (w_{0-})^*  & (f_{0-}^{0+})^* &  (w_{--})^* & M_+^{(-)}   \\
\end{array}%
\right)
\end{equation}
\begin{align}
M_\pm^{(+)} &= K_\pm + \bar{U} (1+|\zeta_+|^2)+
\tilde{U}(2|\zeta_+|^2
+|\zeta_0|^2 -|\zeta_-|^2) + \delta_+,\\
M_\pm^{(0)} &= K_\pm + \bar{U} (1+|\zeta_0|^2)+
\tilde{U}(|\zeta_+|^2
+|\zeta_-|^2) ,\\
M_\pm^{(-)} &= K_\pm + \bar{U} (1+|\zeta_-|^2)+
\tilde{U}(-|\zeta_+|^2
+|\zeta_0|^2 +2|\zeta_-|^2) + \delta_-,\\
f_{kl}^{mn} &= (\bar{U}+\tilde{U})\zeta_k^*\zeta_l+2\tilde{U}
\zeta_m\zeta_n^*,\\
g_{kl}^{m} &= (\tilde{U}-\bar{U})\zeta_k\zeta_l-\tilde{U}
\zeta_m^2,\\
h^{lm}_{k} &= -\bar{U}\zeta_k^2 - 2 \tilde{U}
\zeta_l\zeta_m,\\
w_{kl} &= -(\bar{U}+\tilde{U})\zeta_k\zeta_l,\\
r_{kl} &= (\bar{U}-\tilde{U})\zeta_k^*\zeta_l \,.
\end{align}
Here $\bar{U} = n U_0/\hbar$, $\tilde{U} = n U_2/\hbar$, and
\beq K_{\pm} = -2(J/\hbar) \cos(k a)+(4 J/\hbar) \sin^2(q a/2)
\cos(k a) \pm (2 J/\hbar) \sin(q a) \sin(k a)-\mu\,. \eeq
\end{widetext}

\section{Dynamical stability for $k_1 a>\pi/2$,
$k_2a<\pi/2$}\label{appdynamical}

In this section we analyze the dynamical stability of the
two-component BEC system with equal atom currents $J_1\sin(k_1
a)=J_2\sin(k_2 a)$, when $\cos{(k_1a)}$ and $\cos{(k_2a)}$ exhibit
different signs. Setting the atom currents to be equal allows us to
obtain simple analytic expressions for the stability conditions. The
different signs of $\cos{(k_1a)}$ and $\cos{(k_2a)}$ represent the
situation where the velocities of the two BECs are located on the
opposite sides of the deflection point in the ideal, single-particle
BEC excitation spectrum \eqref{ideal} (the effective masses of the
two components exhibit different signs). Without loss of generality
we assume in the following that $\cos{(k_1a)}<0$ and
$\cos{(k_2a)}>0$. The situation where when $\cos{(k_1a)}$ and
$\cos{(k_2a)}$ have the equal sign is covered in Sec~\ref{2exc}.

The analytic result for the normal mode energies is given by
\eq{2becf}. Similarly to the case when $\cos{(k_1a)}$ and
$\cos{(k_2a)}$ have the same sign, the system is always dynamically
unstable if $\omega_{1,q}^2+\omega_{2,q}^2<0$ and we have the same
condition as in \eq{insta}:
\beq
D_{11}+D_{22}<0\,, \label{insta2}
\eeq
where $D_{ij}$ is defined in \eq{Dco}. Since $k_1a>\pi/2$ and
$k_2a<\pi/2$ and $J_1,J_2>0$, the inequality may even be satisfied
for some values for which $U_{11},U_{22}>0$. In this case only one
of the BECs reaches the (single-component) critical velocity $k a
=\pi/2$, destabilizing the entire two-component BEC system.

Next we assume $\omega_{1,q}^2+\omega_{2,q}^2\geq 0$ and find the
additional unstable regions of the parameter space. When
$\cos{(k_1a)}$ and $\cos{(k_2a)}$ have different signs, the
expression inside the inner square root in \eq{2becf} may become
negative, resulting in a dynamical instability. In particular, this
happens at least for some values of $q$, if
\beq
U_{12}^2 >
\xi_1\,,\label{insta3}
\eeq
where
\beq
\xi_1= -{ (D_{11}-D_{22})^2\over 4n_1
n_2J_1\cos{(k_1a)}J_2\cos{(k_2a)}}\,,\label{xi1}
\eeq
where $D_{ij}$ is defined in \eq{Dco}.

Also the expression inside the outer square root may become
negative. If
\beq
n_1U_{11} < n_2U_{22}\, {\rm min} \(\left| {J_2
\cos(k_2a)\over J_1\cos(k_1a)}\right|, \left| {J_1 \cos(k_1a)\over
J_2\cos(k_2a)}\right|\)\,,\label{u11a}
\eeq
this happens at least for some values of $q$, if
$U_{11}U_{22}>U_{12}^2$. Combining this with \eq{insta3} we find
that the system is {\it stable} for the values of $U_{11}$ that
satisfy \eq{u11a}, if
\beq
U_{11}U_{22}<U_{12}^2<
\xi_1\,.\label{insta4}
\eeq

Similarly, for the values of $U_{11}$ satisfying
\begin{align}
&n_1U_{11}>n_2U_{22} \left| {J_1 \cos(k_1a)\over
J_2\cos(k_2a)}\right| - 4 J_1
\cos(k_1a) \nonumber\\
&n_1U_{11}< n_2U_{22} \left| {J_2 \cos(k_2a)\over
J_1\cos(k_1a)}\right|\,,\label{u11b}
\end{align}
we find that a dynamically {\it stable} system exists if
\beq
\xi_2<
U_{12}^2 < \xi_1\,,\label{insta5}
\eeq
where
\begin{align}
\xi_2 =& U_{11}U_{22} + 4J_1\cos{(k_1a)}J_2\cos{(k_2a)}/(n_1n_2)\nonumber\\
&+ 2[J_1\cos(k_1a)U_{22}/n_1+J_2\cos(k_2a)U_{11}/n_2]\,.\label{xi2}
\end{align}
When $U_{11}$ satisfies \eq{u11b}, $\xi_2$ is always larger than
$U_{11}U_{22}$ and for the dynamically stable region we have
$U_{11}U_{22}<\xi_2<U_{12}^2 < \xi_1$. If $\xi_2>\xi_1$, no stable
region exists.

Note that the {\it entire} stable region in both Eqs.~\eqref{insta4}
and~\eqref{insta5} correspond to the values of the nonlinearities
satisfying $U_{11}U_{22}<U_{12}^2$ that is normally associated with
the dynamically unstable phase separation condition.

The two-component system is therefore dynamically stable if
$D_{11}+D_{22}>0$ and $U_{12}$ satisfies either \eq{insta4} or
\eq{insta5}, for $U_{11}$ defined by \eq{u11a} or \eq{u11b},
respectively. Interestingly, we find a regime where the other
condensate component can stabilize the superfluid flow of an
otherwise unstable condensate. The inequalities \eqref{u11a} or
\eqref{u11b} can be satisfied for $U_{11},U_{22}>0$ when the
component $\psi_1$ exceeds the critical velocity of the
single-component BEC, with $k_1 a>\pi/2$, so that $\omega_{1,q}^2<0$
in \eq{1bec}. The two-component BEC dynamics, nevertheless, is
stable if $U_{12}$ satisfies either \eq{insta4} or \eq{insta5},
respectively.

Note also that we may have, e.g., $U_{11}<0$, $U_{22}>0$, but
$\omega_{1,q}^2+\omega_{2,q}^2>0$ (i.e., $D_{11}+D_{22}>0$). Such a
two-component system can be dynamically stable since
$U_{12}^2>U_{11}U_{22}$.


\begin{thebibliography}{99}
\bibitem{comment} In principle, an external magnetic field may also
induce a different hopping amplitude for different internal levels,
but in the present study we ignore such effects.
\bibitem{comment2} The symmetry of the ground state manifold of the polar
state \eq{polarSoln} in the absence of the magnetic splitting,
$S^1\times S^2/Z_2$ (where $Z_2$ denotes a two-element group), is
reduced in \eq{polar2} to $S^1\times S^1/Z_2$.
\end{thebibliography}
\end{document}